# Features of international taxation and its impact on business entities of Georgia

George Abuselidze and Mariam Msakhuradze

Batumi Shota Rustaveli State University

## Abstract

The work "International Taxation and its impact on Georgian Business Subjects" discusses the essence, types of international taxation and ways to prevent it. Object of international taxation, taxable base and rates, features based on the taxpayer. The approaches of states and its impact on the activities of business entities.

The aim of the work was to study the theoretical and methodological bases of international taxation in the tax system of Georgia and to present the existing problems.

To get acquainted with the activities of the free industrial zones in our country and to evaluate them. Sharing opinions and expressing one's attitude towards it.

The work presents the opinion on the impact of the approaches and recommendations of our country's legislation on international taxation on the business sector of Georgia to correct the current situation.

## ანოტაცია

ნაშრომში „საერთაშორისო დაბეგვრა და მისი გავლენა საქართველოს ბიზნეს-სუბიექტებზე" საკვანძო საკითხებად განხილულია საერთაშორისო დაბეგვრის არსი, სახეები და მისი თავიდან აცილების გზები. საერთაშორისო დაბეგვრის ობიექტი, დასაბეგრი ბაზა და განაკვეთები, თავისებურებები გადასახადის გადამხდელიდან გამომდინარე. სახელმწიფოების მიდგომები და მისი გავლენა ბიზნეს-სუბიექტების საქმიანობაზე.

ნაშრომის შესრულება მიზნად ისახავდა საქართველოს საგადასახადო სისტემაში საერთაშორისო დაბეგვრის თეორიულ-მეთოდოლოგიური საფუძვლების შესწავლასა და არსებული პრობლემების წარმოჩენას. ჩვენს ქვეყანაში არსებული თავისუფალი ინდუსტრიული ზონების საქმიანობის გაცნობასა და მათ შეფასებას. მოსაზრებების გაზიარებას და საკუთარი დამოკიდებულების წარმოჩენას აღნიშნულთან დაკავშირებით.

დასკვნის სახით ნაშრომში წარმოდგენილია მოსაზრება იმის შესახებ თუ რა გავლენას ახდენს საქართველოს ბიზნეს სექტორზე საერთაშორისო დაბეგვრასთან დაკავშირებული ჩვენი ქვეყნის კანონმდებლობის მიდგომები და რეკომენდაციები არსებული მდგომარეობის გამოსასწორებლად.

## შესავალი

სახელმწიფოთა შორის ეკონომიკური ურთიერთობები დღითიდღე ვითარდება, აქედან გამომდინარე, კაპიტალის მოსრაობასთან დაკავშირებული დაბეგვრის საკითხები მნიშვნელოვანი ხდება. არც საქართველოა გამონაკლისი ამ კუთხით. ქვეყანა ცდილობს სრულყოს ორმაგი დაბეგვრის თავიდან აცილების მეთოდები და მექანიზმები. გაფორმებული ხელშეკრულებებით სხვა ქვეყნებთან ცდილობს შექმნას საინვესტიციოდ მიმზიდველი გარემო, კონტროლი გაუწიოს არარეზიდენტების საქმიანობას ჩვენს ქვეყანაში, რომლითაც უზრუნველყოფს, როგორც სხვადასხვა სოციალური თუ ეკონომიკური პრობლემის გადაჭრას, ასევე ბიუჯეტში ფულადი რესურსების მობილიზებას.

აღნიშნული პროცესების განხორციელება სახელმწიფოს მხრიდან არ არის მართივი ისე რომ არ დაზიანდეს ადგილობრივი ბიზნესი. ამ მხრივ გვაქვს გარკვეული წინაღმდეგობები. სწორედ ამან გამოიწვია დაინტერესება და ნაშრომის შესრულება ამ მიმართულებით.

კვლევის მიზანი: საქართველოში არსებულ ბიზნეს-სუბიექტებზე საერთაშორისო ორმაგი დაბეგვრის გავლენის წარმოჩენა, რაც გულისხმობს უპირატესობებისა და ნაკლოვანებების ასახვას და რეკომენდაციების შემუშავებას აღნიშნულთან დაკავშირებით.

ძირითადი ამოცანები:

- საერთაშორისო ორმაგი დაბეგვრის არსის, მიზნებისა და ამოცანების შესწავლა;
- საერთაშორისო ორმაგი დაბეგვრის წარმოქნის თეორიულ-მეთოდოლოგიური საფუძვლების შესწავლა;
- საქართველოში ბიზნეს-სუბიექტების კონკურენტული გარემოს დახასიათება და ანალიზი;
- საქართველოში არარეზიდენტი პირების შემოსავლების დაბეგვრის თავისებურებების შესწავლა;
- საქართველოს ბიზნეს-სუბიექტებზე საერთაშორისო დაბეგვრის არსებული პრინციპების გავლენის განხილვა;
- უცხოური გამოცდილების გაცნობა და შეფასება ადგილობრივთან მიმართებაში.

კვლევის ობიექტი: საერთაშორისო ორმაგი დაბეგვრის თავიდან აცილების შესახებ არსებული კანონმდებლობის საფუძველზე საქართველოს ბიზნეს-გარემოში მიმდინარე პროცესები.

კვლევის საგანი: საერთაშორისო ორმაგი დაბეგვრის თავიდან აცილებასთან დაკავშირებული საკითხების პრაქტიკული ანალიზი.

კვლევის თეორიული და მეთოდოლოგიური საფუძველი: ქართველ და უცხოელ ეკონომისტთა ნაშრომები სადაც განხილულია საერთაშრისო დაბეგვრის არსი და მასთან დაკავშირებული პრობლემები, საქართველოს ნორმატიული და საკანონმდებლო აქტები, საერთაშორისო შეთანხმებები, საერთაშორისო ორგანიზაციათა კვლევები, სხვა სახელმწიფო ორგანოთა სტატისტიკური მონაცემების ბაზა, ექსპერტთა მოსაზრებები.

ნაშრომის სტრუქტურა და მოცულობა: სამაგისტრო ნაშრომი მოიცავს კომპიუტერულ ნაბეჭდ 55 გვერდს, შედგება შესავლის, სამი თავის, თითოეულში ორ-ორი პარაგრაფის და დასკვნისაგან. ნაშრომს ერთვის გამოყენებული ლიტერატურისა და წყაროების ჩამონათვალი.

1. საერთაშორისო დაბეგვრის თეორიულ-მეთოდოლოგიური საფუძვლები

1.1 საერთაშორისო დაბეგვრის არსი და სახეები დღესდღეობით, ქვეყნების სოციალურ-ეკონომიკური განვითარება პირდაპირ კავშირშია რეგიონთაშორისი ურთიერთობების გაღრმავებაზე, რაც ხელს უწყობს მათ შორის კაპიტალის გადაადგილებას. მსოფლიოში მიმდინარე გლობალური პროცესები ხელს უწყობს კომპანიებს, ბიზნესეს საქმიანობა აწარმოონ ერთზე მეტ ქვეყანაში. ამ ყველაფრის შედეგია უცხოური კაპიტალის მოძრაობა, რაც მნიშვნელოვან გავლენას ახდენს რეგიონის განვითარებაზე, კონკურენციის ზრდაზე, ტექნოლოგიური თუ ინფრასტრუქტურული პროექტების განხორციელებაზე და ა.შ.

ბუნებრივია, არსებულ სიტუაციაში მნიშვნელოვანია რეგიონების სოციალურ-ეკონომიკური მდგომარეობის ანალიზი, მისი თავისებურებები, განსაკუთრებით კი ისეთი ეკონომიკური სუბიექტის სიდრმისეული შესწავლა როგორიცაა დაბეგვრა.

კომპანიები, რომლებიც რამდენიმე ქვეყანაში აწარმოებენ საქმიანობას, დაინტერესებული არიან მათი საგადასახადო იურისდიქციით, რადგან ისინი ამ იურისდიქციის შესაბამისად არიან გადასახადის გადამხდელები და აქვთ ვალდებულება მათი გადახდისა, საგადასახადო კანონმდებლობის დაცვით.

ეს ყველაფერი მნიშვნელოვან გავლენას ახდენს მათ გადაწყვეტილებებზე, კერძოდ, უდირთ თუ არა მათ კონკრეტულ ქვეყანაში დაიწყონ ბიზნეს საქმიანობა, რამდენად მძიმედ დააწვება მათ ამ ქვეყანაში არსებული საგადასახადო ტვირთი. ზოგიერთი ქვეყანა ცდილობს შექმნას ინვესტიციებისთვის მიმზიდველი გარემო და მიდის გარკვეულ დათმობებზე, აწესებენ შეღავათებს, რათა მოიზიდონ უცხოური კაპიტალი და ხელი შეუწყონ დასაქმების ზრდას.

სახელმწიფოს უფლება აქვს თავისი სოციალურ-ეკონომიკური მდგომარეობიდან გამომდინარე შექმნას საგადასახადო გარემო, დააწესოს გადასახადები და მათი ნორმები, ისე რომ ხელი შეუწყოს თავისი მიზნებისა და ფუნქციების შესასრულებლად საჭირო თანხების აკუმულირებას ბიუჯეტში.

არარეზიდენტი პირი ექცევა როგორც უცხო ქვეყნის საგადასახადო კანონმდებლობის ქვეშ, ასევე რეზიდენტი ქვეყნის წინაშეა ვალდებული გადაიხადოს გადასახადები შემოსავლიდან. აქედან გამომდინარე, ამ კომპანიის საქმიანობიდან მიღებულ შემოსავალზე, გადასახადების ამოღების სურვილი უჩნდება როგორც შემოსავლის წყაროს ქვეყანას, ასევე რეზიდენტ ქვეყანას. აქედან იკვეთება სამმაგი ინტერესი, შემოსავლის წყაროს ქვეყნის, რეზიდენტი ქვეყნისა და გადასახადის გადამხდელის, რომლის შემოსავალი ერთი და იგივე პერიოდში ექვემდებარება ორჯერ დაბეგვრას. ანუ ადგილი აქვს ორმაგ დაბეგვრას.

ორმაგი დაბეგვრა გულისხმობს გადასახადის გადახდას ორჯერ, ისეთ შემოსავალზე, რომელიც მიიღება ერთი და იგივე წყაროდან. ორმაგი დაბეგვრა შეიძლება იყოს ორი სახის: იურიდიული და ეკონომიკური.

პირველი გულისხმობს დაბეგვრის ისეთ სისტემას, როდესაც ერთი ქვეყნის რეზიდენტ პირს, გააჩნია მუდმივი დაწესებულება მეორე ქვეყანაში და იბეგრება ორჯერ, როგორც არარეზიდენტ ქვეყანაში ისევ რეზიდენტ ქვეყანაში. აქედან გამომდინარე, უკავშირდება მუდმივი დაწესებულების ორჯერ დაბეგვრას. მუდმივი დაწესებულება არ არის დამოუკიდებელი იურიდიული პირი, ის არის რეზიდენტი იურიდიული პირის ფილიალი ან წარმომადგენლობა. ეს იურიდიული პირი კი იბეგრება ორჯერ.

რაც შეეხება ეკონომიკურ ორმაგ დაბეგვრას ის დაბეგვრის ისეთი სისტემაა როცა ერთი ქვეყნის რეზიდენტი პირი მეორე ქვეყანაში აფუძნებს შვილობილ კომპანიას ან ფლობს მეორე ქვეყნის რეზიდენტი პირის წილებს ან აქციებს ან ზოგადად იღებს შემოსავალს და ასეთი შემოსავლები შეიძლება დაიბეგროს ორჯერ: გადახდის წყაროსთან ანუ შემოსავლის მიმღების არარეზიდენტ ქვეყანაში და მეორედ – შემოსავლის მიმღების რეზიდენტ ქვეყანაში. ამ შემთხვევაში, ორმაგი იურიდიული დაბეგვრისგან განსხვავებით შემოსავლის მიმღები და შემოსავლის გამცემი ორი სხვადასხვა იურიდიული პირია. შესაბამისად ადგილი აქვს ორმაგ ეკონომიკურ დაბეგვრას. მაგალითად, ტრანსფერული ფასწარმოქმნები დაკავშირებული შეიძლება იყოს ეკონომიკურ ორმაგ დაბეგვრასთან. ეკონომიკური შიდა (და არა საერთაშორისო) ორმაგი დაბეგვრის მაგალითს მიეკუთვნება მოგების დაბეგვრა მოგების გადასახადით, ხოლო შემდგომ დაბეგრილი მოგების დაბეგვრა ისევ გადასახადით (დივიდენდზე გადასახადი).

იმისათვის რომ თავიდან ავიცილოთ ორმაგი დაბეგვრა არსებობს შემდეგი მეთოდები: ჩათვლის (ანუ მეორენაირად კრედიტის) ან გათავისუფლების. ჩათვლის მეთოდი გამოიყენება მაშინ როცა, ეროვნული კანონმდებლობა გადასახადების გადახდის დროს ითვალისწინებს უცხოეთში გადახდილ გადასახადებს. გათავისუფლების მეთოდი კი ითვალისწინებს რეზიდენტი ქვეყნის კომპანიის მიერ მიღებული შემოსავლის გათავისუფლებას ეროვნული კანონმდებლობით გათვალისწინებული გადასახადებისგან.

ორმაგი დაბეგვრის კონვენციები არ გულისხმობს გადასახადებისგან გათავისუფლებას ან ჩათვლას მხოლოდ ერთი ქვეყნის მიერ. ის ეხება ორივე ქვეყანას, თითოეული მათგანი იდებს ვალდებულობას არ დააბეგროს მეორე ქვეყნის რეზიდენტი თავისი გადასახადებით. ეს ყველაფერი ხელს უწყობს კაპიტალის გადინებას იმ ქვეყანაში სადაც არის დაბალი გადასახადები და ხელსაყრელი საინვესტიციო გარემო.

ქვეყნებმა რომ თავიდან აიცილონ იურიდიული ორმაგი დაბეგვრა ისინი მიმართავენ ჩათლის ან გათავისუფლების მეთოდს თავიანთი რეზიდენტების მიმართ. მაგალითად, ჩვენი ქვეყანა იყენებს ჩათვლის მეთოდს იურიდიული პირებისთვის, ხოლო ფიზიკური პირებისთვის გათავისუფლების მეთოდს. ფიზიკურ პირებს, რომლებიც შემოსავალს ღებულობენ უცხოეთში სრულიად ათავისუფლებს საგადასახადო ვალდებულებებისგან.

თავის მხრივ გათავისუფლების მეთოდი შეიძლება იყო ორგვარი: პოზიტიური და ნეგატიური. გათავისუფლება შეიძლება იყოს პროგრესიით და პროგრესიის გარეშე (რაც ნიშნავს შემოსავლის ზრდის მიხედვით პროგრესიული განაკვეთებით დაბეგვრას). როცა პირი თავისუფლდება პოზიტიური მეთოდით, მას უფლება აქვს რეზიდენტ ქვეყანაში დასაბეგრი შემოსავლიდან გამოქითოს საზღვარგარეთ მიღებული შემოსავალი. თუ მაგალითად, საქართველოს რეზიდენტი მიიღებს დასაბეგრ მოგებას მთელი წლის განმავლობაში 200 ათასი ლარის ოდენობით, აქედან 80 ათასი ლარი კი არის საზღვარგარეთ მიღებული, საქართველოში დასაბეგრი მოგება იქნება 120 ათასი ლარი. ეს ნიშნავს, რომ პროგრესიით პოზიტიური გათავისუფლების დროს, რეზიდენტ ქვეყანაში დასაბეგრ შემოსავალში არ გაითვალისწინება საზღვარგარეთ მიღებული შემოსავალი. თუმცა, ამ მეთოდის გამოყენების დროს, საზღვარგარეთ მიღებული შემოსავალი გამოიყენება პროგრესიული განაკვეთების დადგენის მიზნით. რაც პროგრესიის გარეშე პოზიტიური გათავისუფლების მეთოდის გამოყენების დროს არ გაითვალისწინება.

რაც შეეხება ნეგატიურ გათავისუფლებას, აქ რეზიდენტის დასაბეგრი მოგების განსაზღვრისას გაითვალისწინება პირის მიერ არარეზიდენტ ქვეყანაში მიღებული ზარალი. ამ მეთოდს ზოგიერთი ქვეყანა პოზიტიურ გათავისუფლებასთან ერთად იყენებს. ნეგატიური გათავისუფლების მეთოდიც არსებობს ორი სახის: პროგრესიით და პროგრესიის გარეშე.

პროგრესიის გარეშე ნეგატიური გათავისუფლება გულისხმობს რეზიდენტი პირის ერთობლივი შემოსავლიდან არარეზიდენტ ქვეყანაში მიღებული ზარალის გამოქვითვას. თუ მაგალითად: საქართველოს მოქალაქე წლის განმავლობაში მიიღებს დასაბეგრ მოგებას 100 ათასი ლარის ოდენობით, არარეზიდენტ ქვეყანაში კი ექნება ზარალი 20 ათასი ლარი, მაშინ საქართველოში მისი დასაბეგრი მოგება შეადგენს 80 ათას ლარს.

რაც შეეხება პროგრესიით ნეგატიური გათავისუფლების მეთოდს, ამ შემთხვევაში არარეზიდენტ ქვეყანაში მიღებული ზარალი გამოიყენება პროგრესიული განაკვეთების დადგენის მიზნით.

ზოგადად, საერთაშორისო დაბეგვრა დაკავშირებულია სახელმწიფოების უფლებასთან – თავიანთი ბიუჯეტის სასარგებლოდ დაბეგრონ შემოსავალი ან კაპიტალი ორჯერ. დაბეგვრა შეიძლება დაეფუძნოს ორ ძირითად პრინციპს: მოქალაქეობა ან პირის რეზიდენტობა და შემოსავლის წყარო ანუ ტერიტორიულობის პრინციპი. სწორედ ეს პრინციპები იწვევს ორმაგ დაბეგვრას. მაგალითად, ქვეყანას შეუძლია დაბეგროს უცხო ქვეყნის რეზიდენტი თუკი ის შემოსავალს მიიღებს მის ტერიტორიაზე, ამასთანავე ამ პირის დაბეგვრა შეუძლია იმ ქვეყანას რომლის რეზიდენტიც არის ეს პირი. აქედან გამომდინარე, გადასახადის გადახდის მთავარი პრინციპი, სამართლიანობა, რომ არ დაირღვეს ქვეყნები უნდა შეთანხმდნენ, რომ ერთი და იგივე გადასახადით ორჯერ არ დაბეგრონ ერთი და იგივე დაბეგვრის ობიექტი, იქნება ეს კაპიტალი თუ შემოსავალი. ამ შემთხვევაში გადასახადის გადამხდელისთვის არ აქვს მნიშვნელობა მის შემოსავალს რეზიდენტი ქვეყანა დაბეგრავს თუ არარეზიდენტი, მთავარია ერთჯერადი დაბეგვრის პრინციპი იყოს დაცული.[1]

## 1.2 საერთაშორისო დაბეგვრის ფორმირების თეორიული საფუძვლები

თანამედროვე მსოფლიოში სხვადასხვა სახელმწიფო სუბიექტი განსაზღვრავს საკუთარ უფლებამოსილებებს და გავლენის სფეროებს კანონების მიღებით, როგორც საშინაო, ისე საერთაშორისო დონეზე.

გლობალური ეკონომიკის განვითარება და ეტაპობრივი ინტეგრაცია უზრუნველყოფს ბიზნეს ორგანიზაციებს შესაძლებლობა მისცეს შემოსავალი შექმნან მსოფლიოს სხვადასხვა ქვეყანაში, ხოლო სახელმწიფო საზღვრები ბიზნესისთვის უხილავი ხდება. ამასთან, მსოფლიოს ყველა სახელმწიფო ცდილობს მიიღოს გადასახადები ქვეყნის რეზიდენტებისა და არარეზიდენტების მიერ მიღებული შემოსავლიდან, გააფართოვოს თავისი საგადასახადო იურისდიქცია გადასახადის გადამხდელთა შემოსავალზე. სახელმწიფოს საგადასახადო იურისდიქცია ემყარება პირველ რიგში მის სუვერენიტეტს ეროვნულ ტერიტორიაზე. თითოეულ სახელმწიფოს, თავისი ქვეყნის საზღვრებში, აქვს სრული და განუყოფელი უფლება ჩამოაყალიბოს და გამოიყენოს ნებისმიერი კანონი და ნორმები, რომლებიც განსაზღვრავს მოქალაქეებისა და საწარმოების ქცევის წესებს. საგადასახადო სისტემის სტრუქტურა, ზომები, გადასახადები, მათი ამოღების წესი განისაზღვრება თითოეული სახელმწიფოს მიერ და სავალდებულოა მის მთელ ტერიტორიაზე. ორმაგი დაბეგვრა წარმოიქმნება საგადასახადო იურისდიქციის ჭრილში ინტერესების დაპირისპირების შემთხვევაში.

ორმაგი დაბეგვრის პრობლემა თანამედროვე საერთაშორისო პრაქტიკაში ჩნდება, თუ გადასახადის გადამხდელი, რომელსაც აქვს იურიდიული ურთიერთობა ერთ ქვეყანასთან (მუდმივ საცხოვრებელ ქვეყანაში), შემოსავალს იღებს სხვა ქვეყნის ტერიტორიაზე. შემოსავლის წყაროს ქვეყანაში, როგორც წესი, ვრცელდება თავისი საგადასახადო იურისდიქცია საგადასახადო შემოსავლებზე.

---

ეკონომიკური ურთიერთობების პრინციპიდან გამომდინარე, გადასახადის გადამხდელის მუდმივ საცხოვრებელ ქვეყანას, ამავდროულად, შეუძლია თავისი საგადასახადო იურისდიქციის გაფართოება გადასახადის გადამხდელის შემოსავალზე, სამართლებრივი ურთიერთობის პრინციპის საფუძველზე. ორივე ქვეყნის სამართლებრივი ურთიერთობებისა და ეკონომიკური ურთიერთობების პრინციპების ერთდროულად გამოყენების შედეგი შეიძლება იყოს შემოსავლის ორმაგი დაბეგვრა - შემოსავლის წყაროს ქვეყანაში და მუდმივი საცხოვრებელი ადგილის ქვეყანაში.

უფრო მეტიც, გადასახადის გადამხდელი, რეზიდენტობის პრინციპებიდან გამომდინარე, შეიძლება დაიბეგროს ერთდროულად ორ ქვეყანაში, თუ ამ ქვეყნების კანონმდებლობა იყენებს სხვადასხვა კრიტერიუმს რეზიდენტობის სტატუსის დასადგენად. ამ შემთხვევაში, თუ ორივე სახელმწიფო გადასახადს უწესებს თავის რეზიდენტებს მიუხედავად იმისა თუ რომელი ქვეყნის ტერიტორიაზეა ეს შემოსავალი მიღებული, გადასახადის გადამხდელი თავის ყველა შემოსავალზე ორმაგ საგადასახადო ტვირთს ეწევა. ორმაგი იურიდიული დაბეგვრა ხდება იმ შემთხვევაში, თუ ერთი და იგივე გადასახადის გადამხდელი იბეგრება შესაბამისი გადასახადებით იმავე პერიოდის განმავლობაში, ორ ან მეტ ქვეყანაში იმავე დასაბეგრი ობიექტის მიმართ.

ორმაგი იურიდიული დაბეგვრა შეიძლება მოხდეს იმ შემთხვევებში, როდესაც:
• რამდენიმე სახელმწიფოს ეროვნული კანონმდებლობის შესაბამისად გადასახადის გადამხდელი აღიარებულია რეზიდენტად და შესაბამისად, ეკისრება შეუზღუდავი საგადასახადო ვალდებულება დასაბეგრი ობიექტების მიმართ;
• ერთი სახელმწიფოს რეზიდენტს აქვს სხვა ქვეყნის ტერიტორიაზე დაბეგვრის ობიექტი, და ამ ორივე სახელმწიფოს მიერ ეკისრება გადასახადი ამ საგადასახადო ობიექტზე;

ორმაგი დაბეგვრის პრობლემა, შეიძლება ასევე წარმოიშვას საგადასახადო ბაზის არჩევისას და დასაბეგრი შემოსავლის განსაზღვრის წესებში განსხვავებების გამო. საერთაშორისო ორმაგი დაბეგვრის წარმოქმნა ასევე შესაძლებელია გადასახადის გადამხდელის სტატუსში არსებული განსხვავებების გამო: მაგალითად, ერთ ქვეყანაში, საშემოსავლო გადასახადი იწარმოება პარტნიორობაზე, როგორც იურიდიულ პირზე, ხოლო მეორეში, პირდაპირ პარტნიორების შემოსავალზე.

დღეს მსოფლიოს მრავალი ქვეყანა მივიდა დასკვნამდე, რომ ორმაგი დაბეგვრა წარმოადგენს დაბრკოლებას საერთაშორისო ვაჭრობაზე, კაპიტალის თავისუფალ გადაადგილებასა და საერთაშორისო ბაზრების განვითარებაზე. ორმაგი დაბეგვრის აღმოფხვრის ან მისი ტვირთის შემცირების ორი გზა არსებობს: პირველი არის ქვეყნის მხრიდან შიდა საკანონმდებლო ზომების ცალმხრივი მიღება, მეორე - ორმაგი დაბეგვრის პრობლემის რეგულირება საერთაშორისო ხელშეკრულებების გაფორმებით. ორმაგი ეკონომიკური დაბეგვრის აღმოფხვრის საკითხებს, როგორც წესი, წყვეტს ქვეყნების ეროვნული კანონმდებლობა; ამასთან, ზოგიერთ საერთაშორისო საგადასახადო ხელშეკრულებაში ორმაგი ეკონომიკური დაბეგვრის ცალკეული შემთხვევები რეგულირდება ორმაგ საერთაშორისო დაბეგვრის შემთხვევებთან ერთად. პრაქტიკაში მრავალი ქვეყანა აერთიანებს ორივე მიმართულებას, რომლებიც ავსებენ ერთმანეთს და ამავე დროს არ შეიძლება მთლიანად ურთიერთშემცვლელნი იყვნენ.

ხშირად ეროვნული კანონმდებლობით შემოთავაზებული ცალმხრივი აღმოფხვრის მეთოდები (სრული ან ნაწილობრივი გამონაკლისი, საგადასახადო კრედიტი) ემთხვევა საერთაშორისო ხელშეკრულებებით გათვალისწინებულ აღმოფხვრის მექანიზმს. ამასთან, არცერთ ქვეყანას არ შეუძლია სრულად გადაჭრას ორმაგი დაბეგვრის აღმოფხვრის პრობლემა ცალმხრივად, რადგან ნებისმიერ სახელმწიფოს ყოველთვის აქვს ორმაგი ამოცანა: ერთი მხრივ, საბიუჯეტო შემოსავლების საკმარისი დონის უზრუნველსაყოფად, და მეორეს მხრივ, ეკონომიკური განვითარების სტიმულირებისთვის ოპტიმალური პირობების შექმნა (განსაკუთრებით კაპიტალის ნაკადის სფეროში).

ეროვნული კანონმდებლობით საგადასახადო შეღავათების უზრუნველყოფა უფრო მეტად არის ორიენტირებული ამ ქვეყნის გადამხდელთა - ამ ქვეყნის მაცხოვრებლებისთვის ხელსაყრელი პირობების შექმნაზე, ხოლო უცხოური კომპანიებისა და მოქალაქეების საგადასახადო შეღავათების გაფართოებას ყოველთვის აქვს კონკრეტული გეოგრაფიული ადგილმდებარეობა და განისაზღვრება ორმხრივი სავაჭრო, ეკონომიკური, ფინანსური და სხვა ურთიერთობების განვითარების ხარისხით.

ტრადიციულად, გადასახადის გადამხდელის მუდმივ საცხოვრებელ ქვეყანას აქვს ინიციატივა ორმაგი დაბეგვრის პრობლემის გადასაჭრელად. ჩვეულებრივ, გადასახადის გადამხდელთა მუდმივ საცხოვრებელ ქვეყნებში არ ვრცელდება მათი საგადასახადო იურისდიქცია უცხოური წყაროებიდან მიღებული შემოსავლის ნაწილზე, ტერიტორიული საგადასახადო სისტემის ან საგადასახადო კრედიტის სისტემის გამოყენებით.

ქვეყნების უმეტესობა, მიუხედავად ეკონომიკური განსხვავებისა, ფინანსური პოლიტიკის შემუშავებისა, მიზნებისა და პრიორიტეტების, საგადასახადო სისტემებში აერთიანებს ბინადრობის პრინციპსა (ე.ი. ამ ქვეყნებში მუდმივი საცხოვრებელი ადგილის მქონე პირთა დაბეგვრა ყველა შემოსავლისთვის, მათ შორის საზღვარგარეთ მიღებულ ჩათვლით) და ტერიტორიულობის პრინციპს (მაგ. ამ ქვეყნის ტერიტორიაზე მიღებული შემოსავლის გადასახადების დაწესება, მიუხედავად შემოსავლის მიმღებ პირთა მუდმივი ყოფისა).

იმისათვის რომ ქვეყნებმა დაიცვან თავიანთი საგადასახადო უფლებები საგარეო ეკონომიკური საქმიანობის სფეროში, უნდა განსაზღვრონ წესები. მაგალითად:

- გადასახადის გადამხდელთა ფიზიკური და იურიდიული პირების "ეროვნების" დადგენის წესი;

- კომერციული საქმიანობის „ეროვნულობის" განსაზღვრის წესი (ჩვეულებრივ, არსებობს სამი განსხვავებული რეჟიმი: საქმიანობა ბიზნეს ინსტიტუტის მეშვეობით, დამოუკიდებელი აგენტის მეშვეობით და ქვეყანაში ადგილობრივ ფირმებთან გაფორმებული ხელშეკრულებებით);

- შემოსავლის "ეროვნების" (ან შემოსავლის წყაროს) განსაზღვრის წესი. ჩვეულებრივ, მიმართულია მიდგომები: საქონლის რეალიზაციის ან მომსახურების გაწევის ადგილის, გარიგების დადების ადგილის, საქონლის გადატანის ან მომსახურების განხორციელების ადგილის დადგენის და ა.შ.;

- საზღვარგარეთ გადახდილი გადასახადების კომპენსაციის პროცედურა. ფაქტია, რომ, ეროვნული გადასახადის გადამხდელზე, საზღვრებს გარეთ მიღებული შემოსავლის დაბეგვრის შემთხვევაში, ნებისმიერი ქვეყანა იძულებულია განიხილოს სხვა სახელმწიფოს პრიორიტეტული უფლება, გადასახადით დაბეგროს ნებისმიერი საქმიანობა და შემოსავალი, "უცხოელთა" შემოსავალი იურიდიული და ფიზიკური პირებისთვის. ამ უფლების აღიარება შეიძლება იყოს უცხოური გადასახადის გამოქვითვა გადასახადის გადამხდელისათვის გადახდილი გადასახადიდან (ე.წ. საგადასახადო კრედიტის სახით) ან გადასახადის გადამხდელის დასაბეგრი ბაზის დადგენისას უცხოეთში შემოსული გადასახადების უგულებელყოფა, ან უცხოური შემოსავლის სრულად აღმოფხვრა გადასახადის გადამხდელთა საგადასახადო ვალდებულებებისაგან.

- რეგულირების წესები საგადასახადო მიზნებისათვის ე.წ. ტრანსფერტული ფასები. რასაკვირველია, ყველა ზემოთ ჩამოთვლილი პუნქტისთვის, თითოეული ქვეყანა ცდილობს მაქსიმალურად დაიცვას საკუთარი უფლებები, რათა დაიცვას თავისი ფინანსური ინტერესები და უზრუნველყოს სხვა ქვეყნის საგადასახადო ორგანოებთან სადავო სიტუაციების მოსაგვარებლად მყარი საფუძველი.

ეს უფლებები ფორმულირებულია და დაცულია როგორც ჩვეულებრივი საგადასახადო კანონებში ასევე საგადასახადო კოდექსში. სხვადასხვა სახელმწიფოს საგადასახადო კანონების გამოყენებასთან დაკავშირებული დავები უნდა მოგვარდეს დროებითი ან მუდმივი, ორმხრივი ან მრავალმხრივი საგადასახადო ხელშეკრულებების გაფორმებით. (Полежарова, 2009: 12-18 ) ისტორიულად, საერთაშორისო საგადასახადო ხელშეკრულებები, როგორც ინსტრუმენტი ქვეყნებს შორის საგადასახადო ურთიერთობის დარეგულირებისთვის, წარმოიშვა ჯერ კიდევ მე-19 საუკუნეში. უძველესი ცნობილი ხელშეკრულება 1843 წლით თარიღდება. იგი გაფორმდა საფრანგეთსა და ბელგიას შორის და ითვალისწინებდა ინფორმაციის გაცვლასა და ადმინისტრაციული დახმარების გაწევას საგადასახადო საკითხებში. პირველი ხელშეკრულებები, როგორც წესი, არეგულირებს შემოსავლის შეზღუდული წრის საგადასახადო რეჟიმს. მომავალში, კაპიტალისტური მართვის განვითარების შედეგად, წარმოიქმნა ბიზნესის ორგანიზაციის ახალი ფორმები - იურიდიული პირი, რომლის შემოსავალი ფიზიკური პირის შემოსავლისგან დამოუკიდებლად განიხილება.

ერთა ლიგის ეგიდით შემუშავდა ორმხრივი საგადასახადო ხელშეკრულებების ორი მოდელი. ეს საქმიანობა გაგრძელდა მეორე მსოფლიო ომის შემდეგ OECD– ის ინიციატივით.

მრავალწლიანი მუშაობა დასრულდა 1963 წელს სამოდელო კონვენციის მიღებით, შემოსავლისა და კაპიტალის ორმაგი დაბეგვრის თავიდან აცილების შესახებ. OECD- ის კონვენციის ბოლო ცვლილებები განხორციელდა 1997 წლის ბოლოს.

კონვენციამ არა მხოლოდ ჩამოაყალიბა ორმაგი დაბეგვრის აღმოფხვრის ძირითადი პრინციპები, არამედ ასევე სტანდარტიზებული იქნა კონცეფციები, ტერმინები და კრიტერიუმები, რომლებიც გამოიყენება მთელ მსოფლიოში სპეციალისტების მიერ. მოდელი არის 30 სტატიის შეთანხმების პროექტი, რომელიც ყველაზე ხშირად გამოიყენება როგორც საბაზო დოკუმენტი, მსოფლიოს ქვეყნების პარტნიორებთან შეთანხმებების გაფორმების მიზნით. მოდელს თან ახლავს სტატიების კომენტარი, რაც მონაწილე მხარეებს საშუალებას აძლევს ცალსახად განმარტონ შეთანხმების მოდელში გამოყენებული ცნებები. შეთანხმების მოდელი და კომენტარები მუდმივად განახლებულია იმისთვის, რომ გაითვალისწინონ ახალი ტიპის კომერციული საქმიანობის დაბეგვრის სპეციფიკა, მაგალითად, ელექტრონული კომერცია და ინტერნეტი, და განსაკუთრებით მსოფლიო ეკონომიკის გლობალიზაცია.

ამერიკის შეერთებული შტატების ხაზინის დეპარტამენტმა 1976 წელს გამოაქვეყნა საკუთარი სამოდელო ხელშეკრულება, რომელსაც აშშ-ს მთავრობა იყენებს როგორც სხვა ქვეყნებთან ხელშეკრულებების გაფორმების საფუძველი. OECD-ს მოდელთან შედარებით, ამერიკული მოდელი უფრო მკაცრია და მკაცრ შეზღუდვებს აწესებს აღესებს სარგებელის გამოყენებასთან დაკავშირებით, ცდილობს „საგადასახადო თაღლითობის" თავიდან აცილებას და ისეთი ფირმების შექმნას შეუშალოს ხელი, რომელთა მიზანია, თავიდან აიცილონ დაბეგვრა. არსებობს გარკვეული განსხვავებები ტერმინთან "მუდმივი დაწესებულება".

აშშ-ს სამოდელო კონვენცია, მისი შემქმნელთა აზრით, უნდა ჩაითვალოს დამხმარე დოკუმენტად, რომელიც ეყრდნობა OECD მოდელის კონვენციას, შეერთებული შტატების გამოცდილების გათვალისწინებით. უფრო მეტიც, ეს არ არის სავალდებულო მოდელი შეერთებულ შტატებში და მხოლოდ მოლაპარაკების დაწყების საფუძველს წარმოადგენს.

გაეროს ინიციატივით, მომზადდა რეკომენდაციები განვითარებულ და განვითარებად ქვეყნებს შორის ორმაგი დაბეგვრის ხელშეკრულებების გაფორმების შესახებ. გაეროს დოკუმენტები უფრო მეტად იცავს განვითარებადი ქვეყნების ინტერესებს, ხოლო OECD-სა და აშშ-ს რეკომენდაციები შედგენილია განვითარებული ქვეყნების ფირმების ინტერესების შესაბამისად. ფუნდამენტური განსხვავება OECD- სა და გაეროს მოდელებს შორის არის ის, რომ OECD-ის მოდელი დაფუძნებულია რეზიდენტების დაბეგვრის პრინციპზე, რომელიც შეიძლება განხორციელდეს მხოლოდ იმ სახელმწიფოებს შორის ურთიერთობებში, რომლებიც თანაბარია როგორც პოლიტიკურად, ასევე ეკონომიკურად. გაეროს მოდელი ახორციელებს დაბეგვრის ტერიტორიის პრინციპს, რადგან ამ მოდელის გამოყენებით განვითარებული ქვეყნები დაინტერესებულნი არიან გადასახადის გადამხდელთა წრის გაფართოებით.

მუდმივი დაწესებულების ძირითად განმარტებას შეიცავს OECD მოდელი. იგი დაფუძნებულია ორ სხვა მოდელზე (გაერო და აშშ). უფრო მეტიც, მათი შემდგენლები აღიარებენ OECD მოდელის უპირატესობას. ალტერნატიული მოდელის კონვენციები განიხილება, როგორც OECD მოდელის კონვენციის ცვალებადობა, სპეციფიკური ეკონომიკური და პოლიტიკური პირობებისადმი ადაპტირებული.

გაეროს სამოდელო კონვენცია OECD მოდელის კონვენციისაგან განსხვავდება საქონლის მიწოდების, სამშენებლო და სამონტაჟო სამუშაოების, მომსახურების მიწოდებისა და დამოკიდებული აგენტის დადგენის საკითხზე მუდმივი წარმომადგენლობის თვალსაზრისით.

ამჟამად OECD და გაეროს მოდელები წარმოადგენს ორმხრივი ორმაგი საგადასახადო ხელშეკრულებების მთავარ ტიპებს. ორმხრივის გარდა, ასევე არსებობს მრავალმხრივი ხელშეკრულებები შემოსავლისა და კაპიტალის დაბეგვრის შესახებ (ჩრდილოეთ ევროპის ქვეყნებს შორის, აფრიკის სახელმწიფოებს შორის; ორი მრავალმხრივი ხელშეკრულება მოქმედებდა CMEA- ში).

ხელშეკრულებების გამოყენების მრავალი თეორიული საკითხი (მაგალითად, "მუდმივი დაწესებულების" კონცეფცია, "პირი, რომელსაც აქვს შემოსავლის რეალური უფლება" და ა.შ.) კვლავაც მწვავე დისკუსიების საგანია OECD ჟურნალის კომიტეტში და საგადასახადო საერთაშორისო ასოციაციის კონფერენციებზე. (Полежарова 2009: 34 ).

## 2. საერთაშორისო დაბეგვრა და ბიზნეს-სუბიექტების კონკურენტული გარემო

*2.1 საქართველოში რეგისტრირებული ბიზნეს-სუბიექტების კონკურენტული გარემო*

ნებისმიერი ქვეყნისთვის ბიზნეს გარემოს განვითარებას დიდი მნიშვნელობა აქვს. ის განსაკუთრებულ როლს ასრულებს ქვეყნის ეკონომიკაში. კერძოდ, სამუშაო ადგილების შექმნაში, ქვეყნის საგადასახადო შემოსავლების ზრდაში, მთლიანი შიდა პროდუქტის ზრდაში, ტექნოლოგიური განვითარების მიღწევაში და ა.შ.

ჩვენი ქვეყნისთვის კონკურენტული ბიზნეს გარემოს შექმნა ერთ-ერთი მნიშნელოვანი გამოწვევაა. ვერ ვიტყვით, რომ ამ მხრივ გვაქვს სახარბიელო მდგომარეობა, თუმცა ქვეყანამ გარკვეული ნაბიჯები გადადგა მდგომარეობის გასაუმჯობესებლად.

კანონის "მეწარმეთა შესახებ" მიხედვით[2], საქართველოში სამეწარმეო საქმიანობად ითვლება მართლზომიერი და არაერთჯერადი საქმიანობა, რომელიც ხორციელდება მოგების მიღების მიზნით, დამოუკიდებლად და ორგანიზებულად. მეწარმე სუბიექტები არიან: ინდივიდუალური მეწარმე, სოლიდარული პასუხისმგებლობის საზოგადოები (სპს), შეზღუდული პასუხისმგებლობის საზოგადოება (შპს), სააქციო საზოგადოები (სს), და კოოპერატივი. მეწარმე სუბიექტების რეგისტრაციას ახორციელებს საქართველოს იუსტიციის სამინისტროს მმართველობის სფეროში მოქმედი საჯარო სამართლის იურიდიული პირი-საჯარო რეესტრის ეროვნული სააგენტო.

დღესდღეობით საქართველო ერთ-ერთი ქვეყანაა, რომელიც ბიზნესის დაწყების სიადვილით ხასიათდება. სულ რამდენიმე წუთში შესაძლებელი ახალი ბიზნეს-სუბიექტის დარეგისტრირება მარტივი პროცედურების საშუალებით და მცირე მოსაკრებლით. ვნახოთ რა მდგომარეობაა ბიზნესის კუთხით საქართველოში ბოლო რამდენიმე წლის განმავლობაში.

---

[2] „საქართველოს კანონი მეწარმეთა შესახებ", https://matsne.gov.ge/ka/document/view/28408?publication=64

ცხრილი N1: ბიზნეს სუბიექტების რეგისტრაციის თვის დინამიკა 2010-2019

| წელი | იფე | იმ | სპს | კ | შპს | სს | კს | მლანად |
|---|---|---|---|---|---|---|---|---|
| 2010 | | 34678 | 17 | 18 | 11213 | 27 | 3 | 45956 |
| 2011 | | 38632 | 8 | 25 | 14011 | 53 | 1 | 52730 |
| 2012 | | 27624 | 3 | 23 | 14943 | 36 | 0 | 42629 |
| 2013 | | 28157 | 10 | 43 | 17701 | 50 | 1 | 45962 |
| 2014 | | 27778 | 5 | 579 | 17553 | 62 | 0 | 137901 |
| 2015 | | 24808 | 7 | 954 | 18416 | 93 | 1 | 44279 |
| 2016 | | 23650 | 2 | 550 | 21259 | 78 | 0 | 45539 |
| 2017 | | 24259 | 15 | 251 | 25445 | 106 | 1 | 50077 |
| 2018 | | 24907 | 31 | 179 | 25144 | 136 | 3 | 50400 |
| 2019 | | 26648 | 18 | 145 | 22527 | 207 | 0 | 49545 |

წყარო: საჯარო რეესტრის ეროვნული სააგენტო როგორც ცხრილიდან ჩანს, ბიზნეს სუბიექტების რეგისტრაციის რიცხვი 2010 წლიდან 2019 წლამდე საკმაოდ დიდ მაჩვენებელს შეადგენს. მათ შორის ყველაზე დიდი წილი უჭირავს ინდივიდუალურ მეწარმესა და შეზღუდული პასუხისმგებლობის საზოგადოების ფორმის მეწარმე სუბიექტებს. ეს მონაცემები მიუთითებს იმაზე რომ ბიზნესის სფეროში იქმნება კონკურენციის პირობები, თუ რამდენად წარმატებულად ეს უკვე ცალკეული განხილვის საგანია.

მაგალითისთვის, უფრო დეტალურად ვნახოთ რა მდგომარეობა იყო 2019 წლის დეკემბრის თვეში.

საჯარო რეესტრის ეროვნული სააგენტოს ოფიციალური მონაცემებით, 2019 წლის დეკემბრის თვეში ქვეყანაში რეგისტრირებული ბიზნეს სუბიექტების რაოდენობამ 4 157 ერთეული შეადგინა, რომელთა შორის 97.1% სამეწარმეო სუბიექტია (ინდივიდუალური მეწარმე, სოლიდარული პასუხისმგებლობის საზოგადოება, შეზღუდული პასუხისმგებლობის საზოგადოება, სააქციო საზოგადოება, კომანდიტური საზოგადოება, კოოპერატივი, უცხოური სამეწარმეო იურიდიული პირის ფილიალი), ხოლო 2.9% - არასამეწარმეო სუბიექტი (არასამეწარმეო იურიდიული პირი, უცხოური არასამეწარმეო იურიდიული პირის ფილიალი, საჯარო სამართლის იურიდიული პირი). ამავე თვეში ბიზნესის რეგისტრაციის თვის ინდექსი (დეკემბერი 2010=100) 0.950 დონეზე დაფიქსირდა.

    2019 წლის დეკემბრის თვეში დაფიქსირდა ზრდა, როგორც წინა თვესთან (3.7%-ით), ისე - წინა წლის იმავე თვესთან (14.9%-ით) შედარებით. 2019 წლის დეკემბრის თვეში წინა თვესთან შედარებით სამეწარმეო სუბიექტების რაოდენობა 3.1%-ით, ხოლო არასამეწარმეო სუბიექტების რაოდენობა 28.4%-ით გაიზარდა. განსხვავებული ტენდენცია დაფიქსირდა წინა წლის იმავე თვესთან შედარებით: სამეწარმეო სუბიექტების რაოდენობა 16.0%-ით გაიზარდა, ხოლო არასამეწარმეო სუბიექტების რაოდენობა 12.9%-ით შემცირდა. ამავე თვეში ბიზნესის რეგისტრაციის თვის ინდექსი (დეკემბერი 2010=100) სამეწარმეო და არასამეწარმეო სუბიექტების მიხედვით შესაბამისად 0.949 და 0.961 დონეებზე დაფიქსირდა. დეკემბრის თვეში ბიზნეს სუბიექტთა უმეტესობა დარეგისტრირდა ინდივიდუალური მეწარმის (2 089 ერთეული, ანუ რეგისტრირებულ ბიზნეს სუბიექტთა 50.3%) და შეზღუდული პასუხისმგებლობის საზოგადოების (1 893 ერთეული, ანუ რეგისტრირებულ ბიზნეს სუბიექტთა 45.5%) სახით. როგორც ჩანს, ბიზნეს სუბიექტთა 95.8% მხოლოდ აღნიშნულ სამართლებრივ სტატუსებს ანიჭებს უპირატესობას. ამავე თვეში ქვეყანაში დარეგისტრირდა: 121 არასამეწარმეო იურიდიული პირი, 23 კოოპერატივი, 16 უცხოური სამეწარმეო იურიდიული პირის ფილიალი, 13 სააქციო საზოგადოება, 1 სოლიდარული პასუხისმგებლობის საზოგადოება და 1 უცხოური არასამეწარმეო იურიდიული პირის ფილიალი. 2019 წლის დეკემბრის თვეში რეგისტრირებული ბიზნეს სუბიექტების რაოდენობის მიხედვით გამოიყო ტერიტორიული ერთეულების ტოპ ათეული და გამოვლინდა აღნიშნული მონაცემების ზრდა/კლების ტენდენცია წინა თვესთან და წინა წლის იმავე თვესთან შედარებით.[3]

---

[3] საჯარო რეესტრის ეროვნული სააგენტო ბიზნეს სუბიექტები, 2019: 1-3

ცხრილი N2: ბიზნეს სუბიექტების რეგისტრაციის დინამიკა 2018-2019 წლებში

| ინდიკატორები | | დეკემბერი | |
|---|---|---|---|
| | | 2018 | 2019' |
| რაოდენობა | | | |
| სამეწარმეო სუბიექტები | ერთეული | 3478 | 4035 |
| არასამეწარმეო სუბიექტები | ერთეული | 140 | 122 |
| მთლიანად, ბიზნეს სუბიექტები | ერთეული | 3618 | 4157 |
| | | | |
| სტრუქტურა | | | |
| სამეწარმეო სუბიექტები | პროცენტი | 96.1 | 97.1 |
| არასამეწარმეო სუბიექტები | პროცენტი | 3.9 | 2.9 |
| მთლიანად, ბიზნეს სუბიექტები | პროცენტი | 100 | 100 |
| | | | |
| ზრდა/კლება წინა წელთან შედარებით | | | |
| სამეწარმეო სუბიექტები | პროცენტი | -3.9 | 16 |
| არასამეწარმეო სუბიექტები | პროცენტი | 28.4 | -12.9 |
| მთლიანად, ბიზნეს სუბიექტები | პროცენტი | -3 | 14.9 |
| | | | |
| ბიზნესის რეგისტრაციის თვის ინდექსი | | | |
| სამეწარმეო სუბიექტები | კოეფიციენტი | 0.818 | 0.949 |
| არასამეწარმეო სუბიექტები | კოეფიციენტი | 1.102 | 0.961 |
| მთლიანად, ბიზნეს სუბიექტები | კოეფიციენტი | 0.827 | 0.95 |
| | | | |
| რაოდენობა | | | |
| ინდივიდუალური მეწარმე | ერთეული | 1672 | 2089 |
| სულიდარული პასუხისმგებლობის საზოგადოება | ერთეული | 2 | 1 |
| კოოპერატივი | ერთეული | 7 | 23 |
| შეზღუდული პასუხისმგებლობის საზოგადოება | ერთეული | 1777 | 1893 |
| სააქციო საზოგადოება | ერთეული | 5 | 13 |
| კომანდიტური საზოგადოება | ერთეული | 0 | 0 |
| უცხოური სამეწარმეო იურიდიული პირის ფილიალი | ერთეული | 15 | 16 |
| არასამეწარმეო იურიდიული პირი | ერთეული | 137 | 121 |
| უცხოური არასამეწარმეო იურიდიული პირის ფილიალი | ერთეული | 0 | 1 |
| საჯარო სამართლის იურიდიული პირი | ერთეული | 3 | 0 |

**წყარო**: საჯარო რეესტრის ეროვნული სააგენტო.

**ცხრილი N3:** ტერიტორიული ერთეულების ტოპ ათეული რეგისტრირებული ბიზნეს სუბიექტების რაოდენობის მიხედვით დეკემბერი 2019.

|   | ტერიტორიული ერთეული | რაოდენობა |
|---|---|---|
| 1 | თბილისი | 1860 |
| 2 | ბათუმი | 299 |
| 3 | ქუთაისი | 155 |
| 4 | რუსთავი | 103 |
| 5 | მარნეულის მუნიციპალიტეტი | 97 |
| 6 | გორის მუნიციპალიტეტი | 96 |
| 7 | ზუგდიდის მუნიციპალიტეტი | 88 |
| 8 | გარდაბნის მუნიციპალიტეტი | 63 |
| 9 | ქობულეთის მუნიციპალიტეტი | 58 |
| 10 | თელავის მუნიციპალიტეტი | 56 |

წყარო: საჯარო რეესტრის ეროვნული სააგენტო:

მოცემული ცხრილიდან ჩანს, რომ ყველაზე ბევრი ბიზნეს სუბიექტი რეგისტრირებულია თბილისში, შედარებით მაღალი მაჩვენებელია ბათუმშიც. მათი რაოდენობა მნიშვნელოვნად აღემატება დანარჩენ მუნიციპალიტეტებში რეგისტრირებულ ბიზნეს სუბიექტებს. ეს ყველაფერი, რა თქმა უნდა, განპირობებულია ტერიტორიული ნიშნითა და მოსახლეობის რაოდენობით არსებული განსხვავებებით, თუმცა ფაქტია რომ რეგიონებში არსებული მდგომარეობა დიდად ჩამოუვარდება აღნიშნულ მაჩვენებლებს. მდგომარეობას ვერ შევაფასებთ დადებითად რადგან მნიშნელოვანია რეგიონების განვითარებაზე ზრუნვა, რათა შეიქმნას სამუშაო ადგილები, ჰქონდეთ საკუთარი საბიუჯეტო შემოსავლები,რათა თავიანთი საჭიროებების უზრუნველყოფა მოახდინონ საკუთარი სახსრებით და ნაკლებად იყვნენ დამოკიდებული ცენტრალური ბიუჯეტიდან მიღებულ დაფინანსებაზე.

საყურადღებოა ასევე თავისუფალი ინდუსტრიული ზონის საწარმოების საქმიანობა ჩვენს ქვეყანაში. საქართველოს საგადასახადო კოდექსის თანახმად, თავისუფალი ინდუსტრიული ზონის საწარმო არის ,,თავისუფალი ინდუსტრიული ზონების შესახებ" საქართველოს კანონის[4] შესაბამისად შექმნილი საწარმო. ეს კანონი ადგენს თავისუფალი ინდუსტრიული ზონების შექმნისა და ლიკვიდაციის წესს, განსაზღვრავს თავისუფალი ინდუსტრიული ზონების მართვისა და ამ ზონების ფარგლებში მმართველობისა და მომსახურების/ზედამხედველობის ორგანოების შექმნისა და საქმიანობის წესს, ადგენს თავისუფალი ინდუსტრიული ზონის საწარმოებისათვის დამატებით პირობებს, საგადასახადო შეღავათებსა და ამ ზონების საქმიანობასთან დაკავშირებულ სხვა საკითხებს. ამ ზონაში განხორციელებული ეკონომიკური საქმიანობა ექვემდებარება სპეციალურ ეკონომიკურ და სამართლებრივ რეჟიმს.

მათ საქმიანობასთან დაკავშირებულ საკითხებს შემდეგ თავში უფრო დეტალურად განვიხილავთ.

## 2.2 საერთაშორისო დაბეგვრის ადმინისტრირების თავისებურებები

ნებისმიერი ქვეყანა თავისი საგადასახადო პოლიტიკიდან გამომდინარე შეიმუშავებს მეთოდებს, რათა თავიდან აიცილოს ორმაგი დაბეგვრა. საერთაშორისო სამართალი არ ითვალისწინებს საერთაშორისო ორმაგი დაბეგვრის ამკრძალავ საერთო წესებს. სახელმწიფოს უფლება აქვს აკრიფოს გადასახადი თავის ტერიტორიაზე ეროვნული საგადასახადო კანონმდებლობის შესაბამისად. ამიტომ, პრაქტიკაში საერთაშორისო ორმაგი დაბეგვრის თავიდან აცილების მიზნით გამოიყენება, როგორც ცალმხრივი ღონისძიებები, რომლებიც დაკავშირებულია შიდასახელმწიფოებრივი საგადასახადო კანონმდებლობის შესაბამის ნორმებთან, ასევე მრავალმხრივი ღონისძიებები, რომლებიც დაკავშირებულია ორმაგი დაბეგვრის თავიდან აცილების შესახებ საერთაშრისო შეთანხმებებთან და კონვენციებთან. საერთაშორისო ორმაგი დაბეგვრის თავიდან აცილების ფორმებია:

- განთავისუფლების მეთოდი;
- ჩათვლის მეთოდი;
- გამოქვითვის მეთოდი.

განთავისუფლების მეთოდის შემთხვევაში სახელმწიფო უარს აცხადებს იმ გადასახადით დასაბეგრი ობიექტის დაბეგვრაზე, რომელიც იბეგრება სხვა სახელმწიფოში. შეთანხმებები, რომლებიც, გამოყენებულია მოცემული მეთოდი, განსაზღვრავს იმ შემოსავლებისა და დაბეგვრის ობიექტების ცალკეულ სახეებს, რომლებიც იბეგრება ერთ ქვეყანაში და თავისუფლდება მეორეში.

---
[4] კანონი თავისუფალი ინდუსტრიული ზონების შესახებ" მუხლი 1.
https://matsne.gov.ge/ka/document/view/21994?publication=8

ჩათვლის მეთოდის შემთხვევაში გადამხდელის დასაბეგრი ბაზა ყალიბდება იმ შემოსავლების გათვალისწინებით, რომლებიც მიღებულია საზღვარგარეთ, ხოლო შემდეგ გადასახადების გამოანგარიშებული თანხებიდან გამოიქვითება საზღვარგარეთ გადახდილი გადასახადები.

გამოქვითვის მეთოდის შემთხვევაში საზღვარგარეთ გადახდილი გადასახადი განიხილება ხარჯის სახით, რომელიც გამოიქვითება დაბეგვრას დაქვემდებარებული შემოსავლების თანხიდან. ვინაიდან გამოქვითვის მეთოდი საზღვარგარეთ გადახდილი გადასახადის თანხით ამცირებს მხოლოდ დასაბეგრ ბაზას და არა თვითონ გადასახადს, მისი გამოყენება ნაკლებად ხელსაყრელია, ვიდრე საგადასახადო ჩათვლის გამოყენება. (ვერულიძე, 2015: 196-199) ვნახოთ რა მდგომარეობაა ამ კუთხით საქართველოში. ჩვენი ქვეყნის საგადასახადო კოდექსის თანახმად[5], 1) საწარმო უფლებამოსილია ჩაითვალოს საქართველოს ფარგლებს გარეთ გადახდილი მოგების გადასახადი შესაბამისი საგადასახადო წლისათვის ამ მოგებაზე საქართველოში გადასახადის გადახდისას იმ შემოსავალზე, რომელიც მიღებული არ არის საქართველოში არსებული წყაროდან. 2) საწარმო, რომელიც მოგების გადასახადით მოგების განაწილებისას იბეგრება, უფლებამოსილია გაცემული დივიდენდის მიხედვით მოგების გადასახადის გადახდისას შესაბამისი საგადასახადო პერიოდისათვის ჩაითვალოს საქართველოს ფარგლების გარეთ გადახდილი მოგების გადასახადი იმ შემოსავალზე, რომელიც მიღებული არ არის საქართველოში არსებული წყაროდან.

3) ჩათვლილი თანხების ოდენობა არ უნდა აღემატებოდეს გადასახადების იმ თანხების ოდენობას, რომლებიც ამ მოგებაზე საქართველოში იქნებოდა დარიცხული საქართველოში არსებული წესითა და განაკვეთებით.

ორმაგი დაბეგვრის თავიდან აცილების შესახებ საერთაშორისო შეთანხმებებით განსაზღვრული საგადასახადო შეღავათით სარგებლობისა და არარეზიდენტისათვის საქართველოში გადახდილი გადასახადის დაბრუნების წესი განისაზღვრება საქართველოს ფინანსთა მინისტრის ბრძანებით.

---

[5] ,,საქართველოს საგადასახადო კოდექსი": მუხლი 124.
https://matsne.gov.ge/ka/document/view/1043717?publication=162

ორმაგი დაბეგვრის თავიდან აცილების მიზნით საქართველოს დღესდღეობით 56 ქვეყანასთან აქვს შეთანხმება გაფორმებული. ამ შეთანხმებათა ტექსტები ეფუძნება OECD-ის მოდელს. მათ საფუძველზე ქვეყანა ღებულობს ვალდებულებას, გადასახადის გადამხდელს ჩაუთვალოს წყაროს ქვეყანაში გადახდილი გადასახადი.

ეს შეთანხმებები ძირითადად ვრცელდება შემდეგ გადასახადებზე: საწარმოთა მოგების გადასახადი, ფიზიკურ პირთა საშემოსავლო გადასახადი, საწარმოთა ქონების გადასახადი, ფიზიკურ პირთა ქონების გადასახადი, კაპიტალის მატებაზე გადასახადი.

## 3. საერთაშორისო დაბეგვრის გავლენა საქართველოში და უცხოური გამოცდილება

### 3.1 საერთაშორისო დაბეგვრის გავლენა ბიზნეს-სუბიექტების საქმიანობაზე

ორმაგი დაბეგვრის თავიდან აცილების სრულყოფის მიზნით მნიშვნელოვანია საქართველოში არარეზიდენტი პირების მიერ მიღებული შემოსავლის დაბეგვრის ფაქტორების ანალიზი. არარეზიდენტი შემოსავალს ღებულობს მუდმივი დაწესებულების ან მის გარეშე მიღებული საქმიანობით. საქართველოს საგადასახადო კოდექსის თანახმად, მუდმივ დაწესებულებად ითვლება ადგილი, რომლის საშუალებითაც ეს პირი ნაწილობრივ ან მთლიანად ახორციელებს ეკონომიკურ საქმიანობას. ასეთ შემოსავლებს მიეკუთვნება: დაქირავებით მიღებული შემოსავალი, საქართველოს ტერიტორიაზე საქონლის მიწოდებით მიღებული შემოსავალი და მომსახურების გაწევით მიღებული შემოსავალი.

არარეზიდენტის მიერ საქართველოში არსებული წყაროდან მიღებული შემოსავალი იბეგრება შემდეგი განაკვეთებით:

-დივიდენდები 5%;

-პროცენტები 20%;

-როიალტი 5 %;

-საწარმოს, ორგანიზაციის ან/და მეწარმე ფიზიკური პირის მიერ საერთაშორისო კავშირგაბმულობის ტელესაკომუნიკაციო მომსახურებისათვის და საერთაშორისო გადაზიდვების სატრანსპორტო მომსახურებისათვის გადახდილი თანხები 10%;

- „ნავთობისა და გაზის შესახებ" საქართველოს კანონით განსაზღვრული ნავთობისა და გაზის ოპერაციების განხორციელებისას არარეზიდენტი ქვეკონტრაქტორების მიერ მიღებული შემოსავალი 4%;

-გადახდილი სხვა თანხები, რომელიც კოდექსის თანხმად ითვლება საქართველოში არსებული წყაროდან მიღებულ შემოსავლად 10%;

-ხელფასის სახით მიღებული შემოსავალი 20%.

საქართველოს საგადასახადო კოდექსის თანახმად[6], არარეზიდენტი პირი, რომელიც შემოსავალს იღებს საქართველოში არსებული წყაროდან ან საქმიანობას ახორციელებს მუდმივი დაწესებულების მეშვეობით, მოგების გადასახადის გადამხდელია. დაბეგვრის ობიექტია ერთობლივი შემოსავალი შემცირებული კოდექსით გათვალისწინებული გამოქვითვების თანხებით.  უცხოური საწარმოები, რომლებიც საქმიანობას ახორციელებენ მუდმივი დაწესებულების გარეშე იბეგრება გადახდის წყაროსთან გამოქვითვების გარეშე.

იმისათვის, რომ შევაფასოთ საქართველოში არარეზიდენტი პირის  დაბეგვრის თავისებურებები, საჭიროა გავაანალიზოთ „ორმაგი დაბეგვრის თავიდან აცილების შესახებ" გაფორმებული ხელშეკრულებები სხვა ქვეყნებთან. როგორც აღვნიშნეთ, საქართველოს 56 ქვეყანასთან აქვს გაფორმებული აღნიშნული შეთანხმებები[7].

საქართველოს საგადასახადო კოდექსის თანახმად, პირი ითვლება საქართველოს რეზიდენტად, თუკი ის ფაქტობრივად იმყოფება საქართველოს ტერიტორიაზე 182 ან მეტი დღეს ნებისმიერი უწყვეტი 12 თვის განმავლობაში.

აღსანიშნავია ის ფაქტი, რომ ამ ქვეყნებში არარეზიდენტების დაბეგვრის რეჟიმები განსხვავებულია, კერძოდ, განსხვავდება მათი საგადასახადო განაკვეთები. ორმაგი დაბეგვრის თავიდან აცილების მექანიზმის შემუშავებისას გასათვალისწინებელია ის ფაქტი, რომ საქართველო იყოს მიმზიდველი ქვეყანა საინვესტიციო კუთხით. უნდა დადგინდეს გადასახადის განაკვეთი იმ დონით რომ მან უზრუნველყოს როგორც ფისკალური ისე მასტიმულირებელი ფუნქცია.

ზოგადად, ორმაგი დაბეგვრის თავიდან აცილების შესახებ გაფორმებული ხელშეკრულების მიზანია ხელი შეუწყოს ქვეყნებს შორის ეკონომიკური ურთიერთობების განვითარებას. მნიშნელოვანია ისიც, რომ აღნიშნული ხელშეკრულებებით გათვალისწინებული უნდა იყოს ორივე მხარის ინტერესები. ეს გამოკვეთილი უნდა იყოს საგადასახადო განაკვეთებში, რათა ქვეყანა იყოს საინვესტიციოდ მიმზიდველი. სწორედ, საქართველოს სჭირდება ხელშეკრულებების საფუძველზე დაბალ საგადასახადო განაკვეთებზე უპირატესობის მინიჭება.

---

[6] საქართველოს საგადასახადო კოდექსი: მუხლი 134.
https://matsne.gov.ge/ka/document/view/1043717?publication=162
[7] ორმაგი დაბეგვრის თავიდან აცილების შესახებ შეთანხმებები
https://mof.ge/5127

ვნახოთ რა მდგომარეობაა ამ კუთხით ჩვენს ქვეყანაში. განვიხილოთ გადასახადის სახეების მიხედვით შედავათები ორმაგი დაბეგვრის ხელშეკრულების შესაბამისად.

ცხრილი N4: გადასახადის სახეების მიხედვით შედავათები ორმაგი დაბეგვრის ხელშეკრულების შესაბამისად

| ქვეყანა | მუდმივი დაწესებულება | დივიდენდები | პროცენტი |
|---|---|---|---|
| ავსტრიის რესპუბლიკა | 6 თვე | 0%/ 5%/ 10% | 0 |
| აზერბაიჯანის რესპუბლიკა | 6 თვე | 0.1 | 0.1 |
| გერმანიის ფედერაციული რესპუბლიკა | 6 თვე | 0%/ 5%/ 10% | 0 |
| დანიის სამეფო | 6 თვე | 0%/ 5%/ 10% | 0 |
| ესპანეთი | 6 თვე | 0%/ 10% | 0 |
| ესტონეთის რესპუბლიკა | 9 თვე | 0 | 0 |
| თურქეთის რესპუბლიკა | 12 თვე | 0.1 | 0.1 |
| იტალიის რესპუბლიკა | 6 თვე | 5%/ 10% | 0 |
| ლიტვის რესპუბლიკა | 9 თვე | 5%/ 15% | 0.1 |
| ლატვიის რესპუბლიკა | 6 თვე | 0.05 | 0.05 |
| საბერძნეთის რესპუბლიკა | 9 თვე | 0.08 | 0.08 |
| საფრანგეთის რესპუბლიკა | 6 თვე | 0%/ 5%/ 10% | 0 |
| შვედეთი | 12 თვე | 0%/ 10% | 0.05 |
| ლიხტენშტაინი | 9 თვე | 0 | 0 |
| მოლდოვა | 12 თვე | 0.05 | 0.05 |
| საუდის არაბეთი | 6 თვე | 5%/ 0% | 5%/ 0% |

წყარო: საქართველოს ფინანსთა და ეკონომიკის სამინისტრო:

მოცემული ცხრილიდან ჩანს, რომ საგადასახადო განაკვეთი საქმიანობის მიმართულებების მიხედვით არის განსხვავებული. გადასახადის განაკვეთი მერყეობს 0-დან 15%-მდე, რაც იძლევა იმის თქმის საფუძველს რომ ქვეყანა ინარჩუნებს დაბალი საგადასახადო განაკვეთების სიდიდეს, რაც ხელშემკვრელი სახელმწიფოებისათვის უნდა იყოს მისადები. ამის თქმის საფუძველს გვაძლევს ანალოგიური განაკვეთების შედარება სხვა ქვეყნებთან. მაგალითად, გერმანიაში არარეზიდენტის მიერ მიღებული დივიდენდი იბეგრება 26,38%-ით, პოლონეთში-19%, იტალიაში-26%.

არარეზიდენტის პირების დაბეგვრის ანალიზის დროს საყურადღებოა შემოსავლების ის ჯგუფი, რომლებიც გაერთიანებულია "სხვა შემოსავლებში", რომელიც ითვალისწინებს, სხვადასხვა საქმიანობიდან მიღებული შემოსავლები, რომლებიც კოდექსის მოთხოვნებიდან გამომდინარე გათავისუფლებულია დაბეგვრისგან. ვფიქრობ, აღნიშნული მუხლი არის საყურადღებო, რადგან ქვეყანამ შესაძლოა ამ წყაროდან მიიღოს მნიშვნელოვანი საბიუჯეტო შემოსავლები.

გლობალიზაციის პირობებში ეკონომიკისთვის განსაკუთრებულ მნიშნელობას იძენს ინვესტიციები. ქვეყნის განვითარებასა და ეკონომიკურ სტაბილურობაში მას მნიშვნელოვანი როლი აკისრია. სწორი საინვესტიციო პოლიტიკის გატარება ხელს უწყობს ისეთი პრობლემების გადაწყვეტას როგორებიცაა: სამუშაო ადგილების შექმნა, ტექნოლოგიური პროცესების განვითარება, მოსახლეობის ცხოვრების დონის ზრდა, კონკურენტული გარემოს შექმნა, ინფრასტრუქტურული პროექტების განხორციელება და სხვა. ინვესტიციების მართვის საკითხი ეკონომიკისთვის ერთ-ერთ მთავარ მიმართულებას წარმოადგენს, რადგან დაცული უნდა იყოს როგორც ადგილობრივი ეკონომიკის ინტერესები ასევე ინვესტორის.

"ინვესტიცია" მრავალი თვალსაზრისით შეიძლება განისაზღვროს სხვადასხვა თეორიებისა და პრინციპების მიხედვით. ეს არის ტერმინი, რომელიც შეიძლება გამოყენებულ იქნას მთელ რიგ კონტექსტებში. საერთოდ, ინვესტიცია არის ფულის გამოყენება უფრო მეტი ფულის გამომუშავებისთვის. ინვესტიცია წარმოადგენს რესურსების გამოყენებას მომავალში შემოსავლის ან წარმოების გამომუშავების მიზნით.

თუ ინვესტიცია ეფექტურია, მან ასევე უნდა გაზარდოს ეკონომიკის პროდუქტიული შესაძლებლობები. მაგალითად, განათლების სფეროში ინვესტიციამ შეიძლება გაზარდოს შრომის ნაყოფიერება. ახალ ტექნოლოგიასა და კაპიტალში ინვესტიციამ შეიძლება გაზარდოს პროდუქტიულობა და ეკონომიკის პროდუქტიული შესაძლებლობები; მას შეუძლია ეკონომიკური ზრდა ინფლაციის გარეშე. თუ ინვესტიცია იწვევს პროდუქტიულობის მნიშვნელოვან ზრდას, მაშინ ამან შეიძლება გამოიწვიოს ეკონომიკური ზრდის გრძელვადიანი ტენდენციის ტემპის ზრდა.

ცხრილი N5: საქართველოში განხორციელებული ინვესტიციების რაოდენობა 2013-2019 წლებში (მლნ აშშ დოლარი)

| წელი | 2013 | 2014 | 2015 | 2016 | 2017 | 2018 | 2019 |
|---|---|---|---|---|---|---|---|
| სულ | 1039.2 | 1837 | 1729.1 | 1650.3 | 1962.6 | 1265.2 | 1267.7 |
| I კვ | 291.8 | 331.9 | 343.4 | 392.2 | 411.7 | 323.5 | 283.6 |
| II კვ | 224.1 | 217.6 | 493.2 | 452.1 | 394.0 | 403.6 | 209.8 |
| III კვ | 271.6 | 749.5 | 531.1 | 506.5 | 627.9 | 367.0 | 427.4 |
| IV კვ | 251.6 | 538.0 | 361.3 | 299.5 | 529.0 | 171.0 | 347.0 |

წყარო: საქართველოს სტატისტიკის ეროვნული სამსახური ბოლო 7 წლის განმავლობაში ყველაზე მეტი ინვესტიცია განხორციელდა 2014 წლის მე-3 კვარტალში, რაც მნიშვნელოვნად აღემატება წინა წლების მაჩვენებლებს. ეკონომიკის სექტორების მიხედვით 2019 წლის მონაცემებით, ყველაზე დიდი წილი მოდის საფინანსო სექტორსა და ენერგეტიკაზე.

დიაგრამა N1: პირდაპირი უცხოური ინვესტიციები ეკონომიკის სექტორის მიხედვით 2019 წელი.

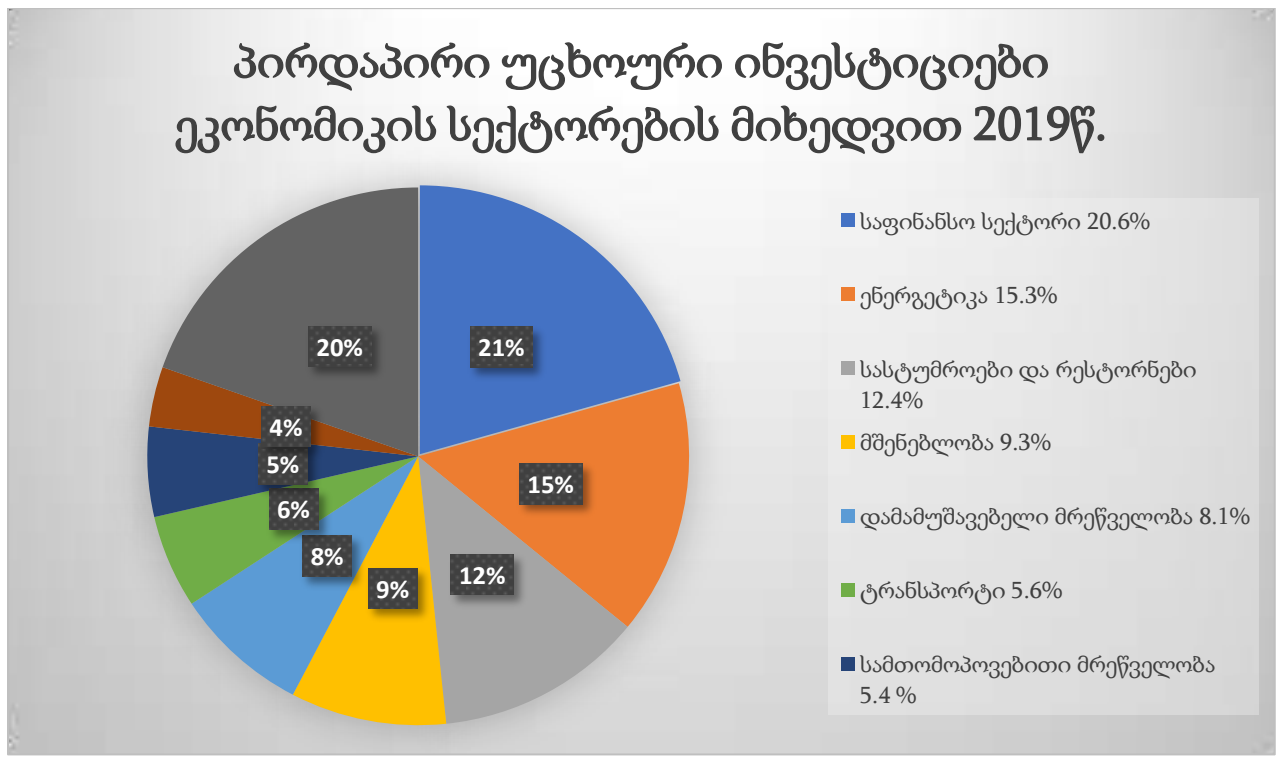

წყარო: საქართველოს სტატისტიკის ეროვნული სამსახური უმსხვილესი ინვესტორების მიხედვით პირველ ადგილზეა გაერთიანებული სამეფო, შემდეგ-თურქეთი.

**დიაგრამა N2:** უმსხვილესი ინვესტორი ქვეყნები 2019 წელი. (მლნ აშშ დოლარი)

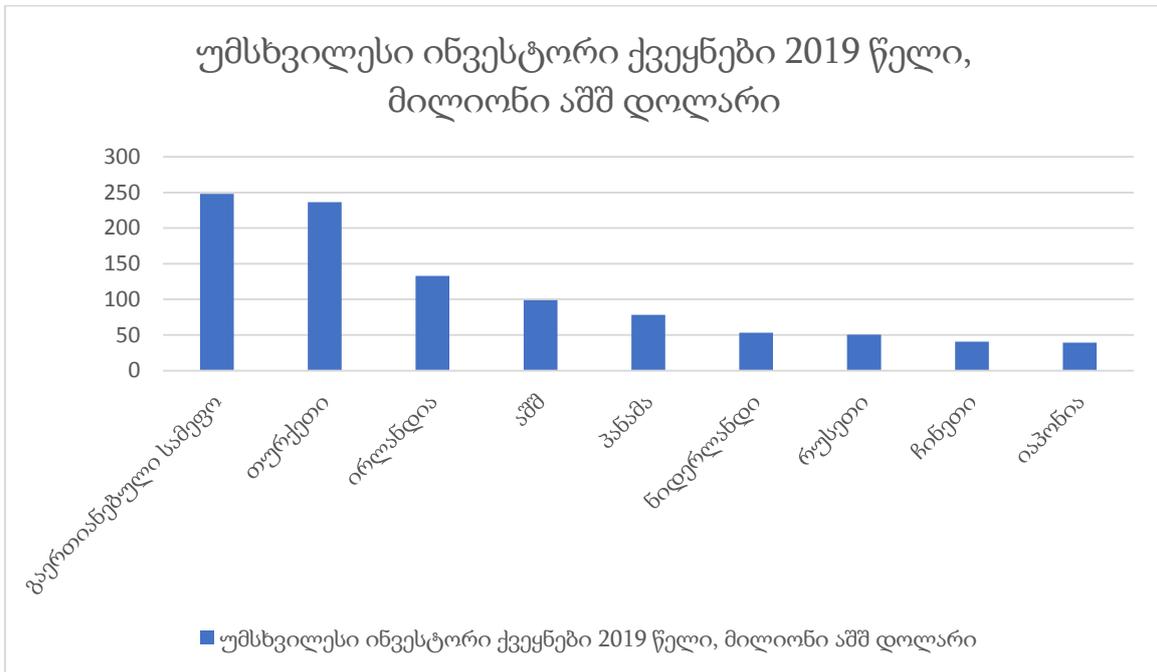

**წყარო: საქართველოს სტატისტიკის ეროვნული სამსახური:**

საქართველოს მიერ ორმაგი დაბეგვრის თავიდან აცილების შესახებ დადებული შეთანხმებებიდან[8] ჩანს რომ ყველაზე მსხვილი ინვესტორები გაერთიანებული სამეფოსა და თურქეთის მიერ მიღებული დივიდენდები იბეგრება ორივე სახელმწიფოში, ასევე განსაზღვრულია რომ, დასაბეგრი თანხა არ უნდა აღემატებოდეს თურქეთის შემთხვევაში დივიდენდის საერთო თანხის 10%-ს, ხოლო გაერთიანებული სამეფოს შემთხვევაში 15%-ს.

მართალია, ზემოთ აღნიშნული ქვეყნები წარმოადგენენ იმ ქვეყნების ნაწილს რომლებიც, აწარმოებენ მსოფლიო შიდა პროდუქტის დიდ ნაწილს და საქართველოს ეკონომიკას მათთან შედარებით გაუჭირდება თანაბარი მიმზიდველი პირობების შეთავაზებაზე, თუმცა ამ კუთხით ჩვენმა ქვეყანამ უნდა გადადგას ნაბიჯები და შეიმუშაოს ისეთი საგადასახადო სისტემა, რომელიც მისაღები იქნება მათთვის.

პირდაპირ უცხოურ ინვესტიციებს ჩვენი ქვეყნისთვის მრავალი დადებითი შედეგის მოტანა შეუძლია, გამომდინარე იქიდან რომ საქართველოს არ გააჩნია საკმარისი ფინანსური რესურსი იმისათვის რომ გამოიყენოს ადგილობრივი საინვესტიციო შესაძლებლობები. ინვესტიციებისთვის ხელსაყრელი გარემოს შექმნაზე სხვადასხვა ფაქტორი ახდენს გავლენას. ეს შეიძლება იყოს გეოგრაფიული მდებარეობა, პოლიტიკური მდგომარეობა, მუშახელის სიიაფე, განათლების დონე, ეკონომიკური მდგომარეობა, ბუნებრივი რესურსები და ა.შ. აღნიშნულ საკითხზე მუშაობს მსოფლიო ბანკის პროექტი Doing Business,[9] რომელიც ითვალისწინებს ბიზნესის რეგულაციების და მათი შესრულების ობიექტურ ზომებს 190 ეკონომიკასა და შერჩეულ ქალაქში, სუბნაციონალურ და რეგიონულ დონეზე. Doing Business პროექტი, რომელიც 2002 წელს დაიწყო, აკვირდება შიდა მცირე და საშუალო კომპანიებს და ზომავს მათზე გამოყენებულ რეგულაციებს. Doing Business- ის მიერ ბიზნესის რეგულირების გარემოების შედარების მიზნით, ყოვლისმომცველი რაოდენობრივი მონაცემების შეგროვებითა და ანალიზით, Doing Business მოუწოდებს ეკონომიკებს კონკურენცია გაუწიონ უფრო ეფექტურ რეგულირებას; ემსახურება კერძო სექტორის მვლევარებისა და თითოეული ეკონომიკის ბიზნეს კლიმატით დაინტერესებულ პირებს.

---
[8] იხ. საქართველოს მიერ გაფორმებული „ორმაგი დაბეგვრის თავიდან აცილების შესახებ" შეთანხმებები: https://mof.ge/5127

გარდა ამისა, Doing Business გთავაზობთ დეტალურ ნაციონალურ კვლევებს, რომლებიც ამომწურავად აღწერს ბიზნესის რეგულაციას და რეფორმას სხვადასხვა ქალაქსა და რეგიონში. ეს კვლევები მოიცავს მონაცემებს ბიზნესის კეთების სიმარტივის შესახებ, აფასებს თითოეულ ადგილს და რეკომენდაციას უწევს რეფორმებს ინდიკატორების თითოეულ სფეროებში შესრულების გასაუმჯობესებლად. არჩეულ ქალაქებს შეუძლიათ შეადარონ თავიანთი ბიზნესის რეგულაციები ეკონომიკაში ან რეგიონში არსებულ სხვა ქალაქებთან და 190 ეკონომიკასთან, რომლებიც Doing Business- მა დაადგინა. ის ქვეყნების შეფასებას ახდენს სხვადასხვა კრიტერიუმით:

- ბიზნესის დაწყება
- მშენებლობის ნებართვა
- ელექტროენერგიით მომარაგება
- ქონების რეგისტრაცია

---
[9] THE WORLD BANK DOING BUSINESS. https://www.doingbusiness.org/en/about-us

- კრედიტის მიღება
- მინორიტარული ინვესტორების დაცვა
- გადასახადების გადახდა
- ვაჭრობა საზღვრებს შორის
- კონტრაქტის აღსრულება
- გადახდისუუნარობა
- შრომის ბაზრის რეგულაცია.

Doing Business 2020-ის მონაცემებით საქართველო საინვესტიციო მიმზიდველობით მე-7 ადგილზეა. უმნიშვნელო მაჩვენებლით ჩამოუვარდება აშშ-ს.

**ცხრილი N 6**: ქვეყნები მსოფლიო რეიტინგში ბიზნესის კეთების სიმარტივის მიხედვით

| Rank | Economy | DB Score |
|---|---|---|
| 1 | New Zealand | 86.8 |
| 2 | Singapore | 86.2 |
| 3 | Hong Kong SAR, China | 85.3 |
| 4 | Denmark | 85.3 |
| 5 | Korea, Rep. | 84 |
| 6 | United States | 84 |
| 7 | Georgia | 83.7 |

**Source:** Doing Business Database

საქართველო პირველ ათეულშია შემდეგი კომპონენტების მიხედვით:

ბიზნესის დაწყება - მე-2 პოზიცია

ქონების რეგისტრაცია- მე-5 პოზიცია

მინორიტარულ ინვესტორთა დაცვა მე-7 პოზიცია.

2020 წელს ჩვენი ქვეყანა ევროპისა და ცენტრალური აზიის ქვეყნებს შორის ლიდერ პოზიციაზეა.

**დიაგრამა N3:** საქართველოს პოზიცია ბიზნესის კეთების კრიტერიუმების მიხედვით;

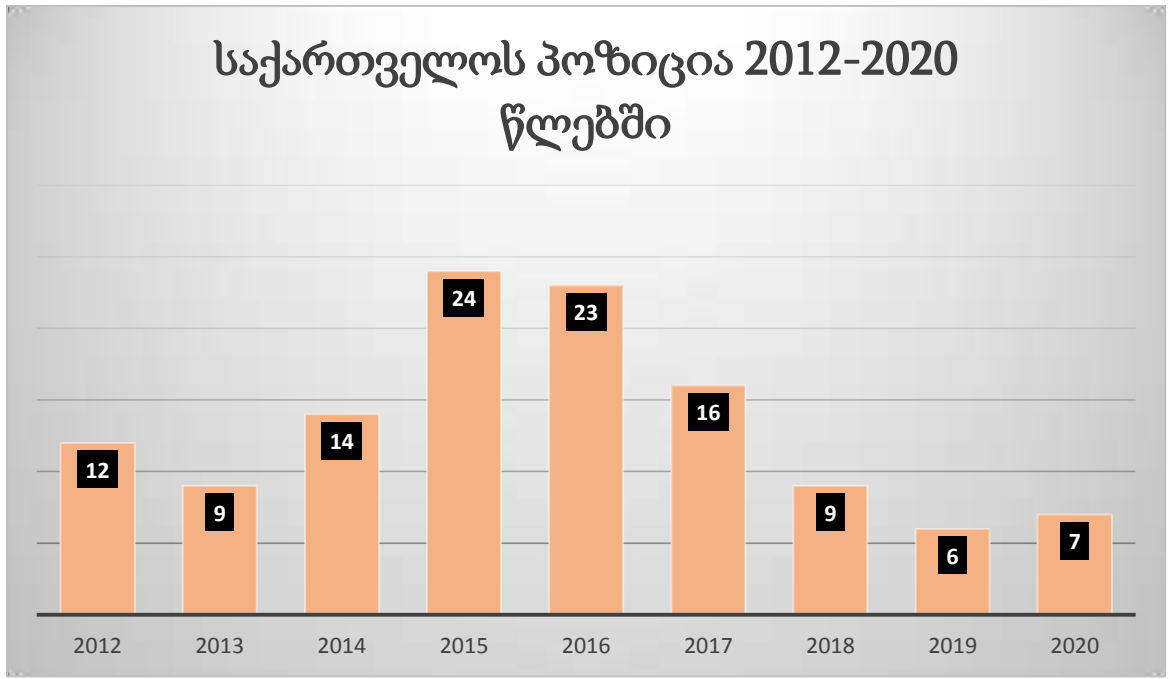

**წყარო:** საქართველოს ეკონომიკისა და მდგრადი განვითარების სამინისტრო ქვეყანაში უცხოური ინვესტიციების შემოდინებაზე გავლენას ახდენს პოლიტიკურ-ეკონომიკური ფაქტორები. განსაკუთრებით, ისეთ ქვეყნებში, რომლებსაც არ აქვთ საკუთარი რესურსები. 2004 წლიდან დღემდე საქართველოში განხორციელდა მნიშვნელოვანი საფინანსო-საგადასახადო რეფორმები[10], რამაც გაამარტივა საგადახადო ადმინისტრირების პროცედურები. პუი-ის მოზიდვის მიზნით, სახელმწიფომ მიიღო კანონი თავისუფალი ინდუსტრიული ზონების შესახებ, რომლის მიხედვითაც ინვესტიციებს მიეცათ გარკვეული შეღავათები. ეს შეღავათები მოიცავდა: გადასახადებისაგან გათავისუფლებას, ექსპორტ-იმპორტის ლიბერალურ პროცედურებს, სტრატეგიულ გეოგრაფიულ ადგილმდებარეობას. შეიქმნა თავისუფალი ინდუსტრიული ზონები თბილისში, ქუთაისსა და ფოთში.

საგადასახადო სფეროში რეფორმების შემუშავების მთავარი მიზანი იყო ინვერსტორებისათვის მასტიმულირებელი გარემოს შექმნა. 2004 წლიდან შემოდებული იქნა სპეციალური საგადასახადო შეღავათები. კერძოდ, დაბეგვრის გამარტივებული პროცედურები, გადასახადების რაოდენობის შემცირება, განსხვავებით სხვა ქვეყნებისგან სოციალური გადასახადის არარსებობა, კაპიტალის დანახარჯის გამოქვითვა ინვესტიციების განხორციელების წელს, ზარალის გადატანა მომდევნო წლებში (საგადასახადო კოდექსის მიხედვით ამ დროისთვის მომდევნო 5 წლით), თავისუფალ ინდუსტრიულ ზონებში გადასახადებისგან გათავისუფლენა საგადასახადო კრედიტის გამოყენებით, ბიზნესის დაწყების პროცედურების გამარტივება, ლიბერალური საგადასახადო განაკვეთები. საგადასახადო კოდექსში შემოდებული იქნა სპეციალური სტატუსები: საერთაშორისო საწარმო და უცხოური საწარმო. უცხოური ინვეტორების მიმართ საგადასახადო პოლიტიკა ასევე ასახულია ხელშეკრულებებში "ორმაგი დაბეგვრის თავიდან აცილების შესახებ", სადაც ასახულია ინვესტორთათვის გარკვეული შეღავათები და გარანტიები. ეს შეღავათები დიფერენცირებულია ქვეყნების მიხედვით და რა თქმა უნდა ლიბერალურია.

---

[10] საგადასახადო პოლიტიკა და პირდაპირი უცხოური ინვესტიციები საქართველოში" ყუფარაძე გ. თბილისი 2013 წ.
https://library.iliauni.edu.ge/wp-content/uploads/2017/03/giorgi-qhupharadze-sagadasakhado-politika-da-pirdapiri-utskhouri-investitsiebi-saqarthveloshi.pdf

საქართველოს საინვესტიციო პირობების შესახებ დადებით შეფასებებს ვხვდებით სხვადასხვა საერთაშორისო რეიტინგში. ასე მაგალითად, ევროპის რეკონსტრუქციისა და განვითარების ბანკის პრეზიდენტი სტატიაში "Georgia Investing For Change" აღნიშნავს, რომ საქართველოში ინვესტიციებმა პოზიტიური ცვლილებები გაიცადა. ქვეყანამ ხელი შეუწყო ინვესტორებს წარმატებული პოლიტიკით და რეფორმებით[11]. ამან განაპირობა ის რომ მათ დაიწყეს იმ ინიციატივების განხორციელება რასაც გეგმავდნენ გარდამავალ ქვეყნებში და საშუალება მიეცათ საქართველოში ინოვაციური საინვესტიციო რესურსების გამოყენებისა. ამიტომ, საქართველოში და EBRD-ის სხვა ქვეყნებში ცდილობენ შექმნან უფრო მძლავრი და მდგრადი ეკონომიკა, რაც ხელს შეუწყობს ინტეგრაციასა და გლობალური ეკონომიკური გამოწვევების გადაჭრას. ეს არის ის სტრატეგია რომელიც დომინირებს მათ საქმიანობაზე მრავალი წლის განმავლობაში.

გამომდინარე იქიდან, რომ ქვეყანას სჭირდება უცხოური ინვესტიციები სახელმწიფო აღნიშნული კუთხით ცდილობს შექმნას ისეთი პირობები, რომლებიც დააინტერესებს უცხოელ ინვესტორებს. ამ მიზნით 2007 წელს საქართველოს მთავრობის ინიციატივით შეიქმნა კანონი "თავისუფალი ინდუსტრიული ზონების შესახებ", რომლის საფუძველზეც ქვეყანაში არის შესაძლებელი თავისუფალი ინდუსტრიული ზონების შექმნა. მოცემული კანონის თანახმად: თავისუფალი ინდუსტრიული ზონა არის ქვეყნის ტერიტორიის ის ნაწილი სადაც მოქმედებს ადგილობრივისაგან განსხვავებული საგადასახადო დაბეგვრის რეჟიმები. ამ ზონას აქვს განსაზღვრული ტერიტორია, რომლის ფართობი აღემატება 10 ჰექტარს, აქვს შესასვლელი და გამოსასვლელი და ეწყობა საბაჟო გამშვები პუნქტები. მისი შექმნა შეიძლება როგორც მთავრობის, ასევე ფიზიკური და იურიდიული პირის ინიციატივით. თავისუფალ ინდუსტრიულ ზონაში შეუძლია საწარმოს დარეგისტრირდეს და მიიღოს ნებისმიერი სამართლებრივ-იურიდიული ფორმა საქართველოს საგადასახადო კანონმდებლობის შესაბამისად. ანგარიშსწორება აღნიშნულ ზონაში შესაძლებელია ნებისმიერი ვალუტით. ისეთ საქმიანობებზე, რომლებიც საჭიროებს ლიცენზიის მიღებას არის გამარტივებული წესები.

---
[11] European Bank for reconstruction and Development- Brochure

თიზ-ში შესაძლებელი ნებისმიერი საქონლის წარმოება და მომსახურების გაწევა, გარდა აკრძალული საქმიანობისა როგორიცაა: იარაღით, საბრძოლო მასალებით, ნარკოტიკული საშუალებებით ვაჭრობა და ა. შ. აქ არ ვრცელდება ადგილობრივი თვითმმართველი ორგანოების უფლებამოსილებები. აღნიშნული ტერიტორია აკრძალულია საცხოვრებელ ადგილად გამოსაყენებლად. თავისუფალ ინდუსტრიულ ზონებში მოქმედებს შემდეგი საგადასახადო შეღავათები:

- თიზ-ის საწარმო გათავისუფლებულია მოგების გადასახადისგან;
- დღგ-თი არ იბეგრება უცხოური საქონლის შეტანა;
- ამ ზონაში განხორციელებული ოპერაციები გათავისუფლებულია დღგ-სგან;
- აქ არსებული ქონება გათავისუფლებულია ქონების გადასახადისგან;
- უცხოური საქონლის შეტანა ასევე გათავისუფლებულია იმპორტის გადასახადისგან;
- თიზ-ში წარმოებული პროდუქციის საქართველოს სხვა ტერიტორიაზე გატანა გათავისუფლებულია იმპორტის გადასახადისგან;
- დაქირავებული საშემოსავლო გადასახადს იხდის დეკლარირების საფუძველზე.

დღესდღეობით ჩვენს ქვეყანაში არსებობს ხუთი თავისუფალი ინდუსტრიული ზონა ესენია: ფოთის თიზ; ყულევის თიზ; ქუთაისის თიზ; თბილისის ტექნოლოგიური პარკის თიზ; ქუთაისის ჰუალინგის თიზ. რამდენიმე სიტყვით დავახასიათოთ თითოეული მათგანი:

**ფოთის თიზ** შეიქმნა 2009 წელს 300 ჰა მიწის ფართობზე, 99 წლის ვადით. მისი ორგანიზატორი იყო არაბული კომპანია „რაკია საქართველო თავისუფალი ინდუსტრიული ზონა". 2016 წელს აღნიშნულმა კომპანიამ თიზი-ს 85% გადასცა სახელწიფოს რადგან მან ვერ შეასრულა ნაკისრი ვალდებულებები და დაეკისრა პირგასამტეხლო. იმისათვის რომ ჩამოეწერა აღნიშნული დავალიანება, ინვესტორმა სახელმწიფოს გადასცა საკონტროლო პაკეტის 85%. 2017 წელს კი ზონის 75% გადაეცა ჩინურ კომპანია China Energy Company Limited-ს, რომელიც ფუნქციონირებს საფინანსო, ენერგეტიკის და სავაჭრო სფეროში. ზონის 10% სახელმწიფოს საკუთრებაში რჩება[12].

**ყულევის თიზ** შეიქმნა 2012 წელს 99 წლის ვადით საქართველოსა და შპს „Socar Georgia Investment"-ს შორის გაფორმებული ხელშეკრულების საფუძველზე. ხოლო 2016 წელს მისი უფლებამონაცვლე გახდა შპს „ფაზის ოილი", რომელმაც ფუნქციონირება დაიწყო 2017 წელს.

**ქუთაისის თიზ** 2009 წელს 27 ჰექტარ მიწის ფართობზე ყოფილი ავტოქარხნის ტერიტორიაზე და ფუნქციონირების ვადა განისაზღვრა 99 წლით. ორგანიზატორი და ადმინისტრატორი არის შპს „ჯორჯიან ინტერნეიშნლ ჰოლდინგი". აღნიშნული ზონა შექმნილია 100%-იანი კერძო კაპიტალით და მის მიზანს წარმოადგენდა როგორც ადგილობრივი ისე უცხოელი ინვესტორების ძალებით მეწარმეობის პრინციპზე აგებულის თიზი-ს ჩამოყალიბებას. ტერიტორიაზე ფუნქციონირებს 37 საწარმო. უმსხვილესი საწარმო არის Fress Georgia LTD, , რომელიც აწარმოებს და რეალიზაციას ახდენს საყოფაცხოვრებო გაზისა და ელექტრო ტექნიკის საქონლის.

**თბილისის ტექნოლოგიური პარკის თიზ**-ს 100 %-იანი მესაკუთრეა „ბითფური ჰოლდინგი ბი ვი", ხოლო ორგანიზატორი შპს „ჯორჯია ტექნოლოჯი პარკი", ის შეიქმნა 2015 წელს და ფუნქციონირების ვადა განისაზღვრა 49წელი. ის მდებარეობს 17 ჰექტარ მიწის ნაკვეთზე, აქვს წვდომა საქართველოს შრომით რესურსზე. დაყოფილია 28 ინდივიდუალურ მიწის ნაკვეთად. უპირატესობით სარგებლობს რადგან არის ახლოს ქალაქის ცენტრათან და საერთაშორისო აეროპორტთან და შეუძლია თვირთის გადაზიდვა მთავარ ავტომაგისტრალთან. ეს ყველაფერი კი იძლევა შესაძლებლობას რომ ჰქონდეს კომუნიკაციებსა და გზებზე წვდომა შეზღუდვების გარეშე. 2015 წელს ბითფური პოლდინგმა გახსნა 40მგვტ სიმძლავრის დატა ცენტრი, რომლის დანახარჯების 90% მოდის ელექტროენერგიაზე.

---
[12] იხ. სტატია „რა აფერხებს საქართველოში თიზების განვითარებას ?" 2017 წელი.
https://forbes.ge/news/3229/ra-aferxebs-Tizebis-ganviTarebas?fbclid=IwAR38IHE7BBtIiEzKMVh08HfirzS9Czd1TaNpYE3I9RknQQvrnPkwut0Y1K0

**ქუთაისის ჰუალინგის თიზ** შეიქმნა 2015 წელს 30 წლის მოქმედების ვადით 36 ჰა მიწის ფართობზე. ორგანიზატორი არის შპს „GEORGIAN HUASHUN INTERNATION INDUSTIAL INVESTMENT GROUP ". აქ ფუნქციონირებს ხუთი საწარმო რომელთა საქმიანობის სფერო არის ავეჯის, ხის, ლითონკონსტრუქციების, სამშენებლო ქვისა და მატრასების წარმოება. [13]

საქართველოს მთავრობა თავისუფალ ინდუსტრიულ ზონებზე დიდ იმედებს ამყარებდა. მათი განცხადებით ამ ზონების მიღების მიზეზი იყო შემდეგი:

- უცხოური კაპიტალის მოზიდვა;

- სამუშაო ადგილების შექმნა;

- რეგიონის ცნობადობის ამაღლება;

- ტექნოლოგიური პროცესების განვითარება;

- ინფრასტრუქტურული პროექტების განხორციელება;

- ექსპორტის სტიმულირება;

- ცოდნის ტრანსფერი ადგილობრივ ეკონომიკაში.

იმასთან დაკავშირებით თუ რა გავლენას ახდენს ეკონომიკაზე თავისუფალი ინდუსტრიული ზონების ფუნქციონირება არსებობს დადებითი და უარყოფითი შეფასებებიც. მთავრობას ჰქონდა მოლოდინი რომ ზონების შექმნა ქვეყნის ეკონომიკას მნიშვნელოვან დადებით ეფექტს მოუტანდა. თუმცა რა მდგომარეობაა ამ მხრივ ქვეყანაში ეს უკვე ცალკე საკითხია. მაგალითისთვის რომ მოვიყვანოთ ფოთის თავისუფალი ინდუსტრიული ზონა, მოლოდინი რომ აქ დუბაი უნდა აშენებულიყო ფაქტია რომ არ არის გამართლებული. იყო მოლოდინი რომ აქ

---

[13] საქართველოს თავისუფალი ინდუსტრიული ზონები, 2018
http://parliament.ge/ge/ajax/downloadFile/98813/%E1%83%A1%E1%83%90%E1%83%A5%E1%83%90%E1%8 3%A0%E1%83%97%E1%83%95%E1%83%94%E1%83%9A%E1%83%9D%E1%83%A1_%E1%83%97%E1%83 %90%E1%83%95%E1%83%98%E1%83%A1%E1%83%A3%E1%83%A4%E1%83%90%E1%83%9A%E1%83% 98_%E1%83%98%E1%83%9C%E1%83%93%E1%83%A3%E1%83%A1%E1%83%A2%E1%83%A0%E1%83% 98%E1%83%A3%E1%83%9A%E1%83%98_%E1%83%96%E1%83%9D%E1%83%9C%E1%83%94%E1%83%9 1%E1%83%98?fbclid=IwAR04WdxIDUg4BHXuZR7jD_zp5apiy-52F3XJfvTggPdctKJHbB2gSRThFos

მოზიდული იქნებოდა 200 მილიონი აშშ დოლარის ინვესტიცია და დასაქმდებოდა ათასობით ადამიანი. მაგრამ, არაბულმა კომპანიამ გაასხვისა ფოთის პორტი და ფოთის თზი-ს დაარსებიდან 4 წლის შემდეგ აქ მხოლოდ ბრინჯის დაფასოების ყაზახურ-უკრაინული კომპანია მუშაობდა სადაც 20-მდე ადამიანი იყო დასაქმებული.

კვლევის პროცესში, როდესაც ვეცნობოდი აღნიშნული ზონების საქმიანობის თავისებურებებსა და სახელმწიფოს მხრიდან მათ მხარდაჭერას, გამიჩნდა უარყოფითი დამოკიდებულება. ჩემი სუბიექტური აზრი დაკავშირებულია როგორც ქვეყნის კანონმდებლობასთან ასევე თავად თავისუფალი ინდუსტრიული ზონების საქმიანობასთან და მათ მიერ ამ შეღავათების სარგებლობასთან თავიანთი ინტერესებიდან გამომდინარე.

პირველ რიგში მინდა აღვნიშნო, რომ ქვეყანას ბიზნესი სჭირდება არამარტო ინფრასტრუქტურული თუ სხვა ამგვარი მიზეზების გამო, არამედ საბიუჯეტო შემოსავლების ზრდისათვის. ვიცით რომ ბიუჯეტის შემოსავლების დაახლოებით 80% სწორედ რომ გადასახადებზე მოდის. საბოლოო ჯამში ბიზნესის ხელშეწყობის მთავარ მოტივს წარმოადგენს დასაბეგრი ბაზის ზრდა, რაც გამოიწვევს საბიუჯეტო შემოსავლების ზრდას. თავისუფალ ინდუსტრიულ ზონებში, როგორც ზემოთ აღვნიშნეთ, კომპანიები გათავისუფლებული არიან გადასახადებისგან. იხდიან მხოლოდ საშემოსავლო გადასახადს. ეს იმას ნიშნავს რომ ქვეყნის შემოსავლების ზრდაზე აღნიშნულ ზონაში მოქმედი საწარმოების საქმიანობა არანაირ დადებით ეფექტს არ იღებს.

დიაგრამა N4: თიზ-ს საწარმოსა და ადგილობრივ საწარმოს შედარებითი ანალიზი.

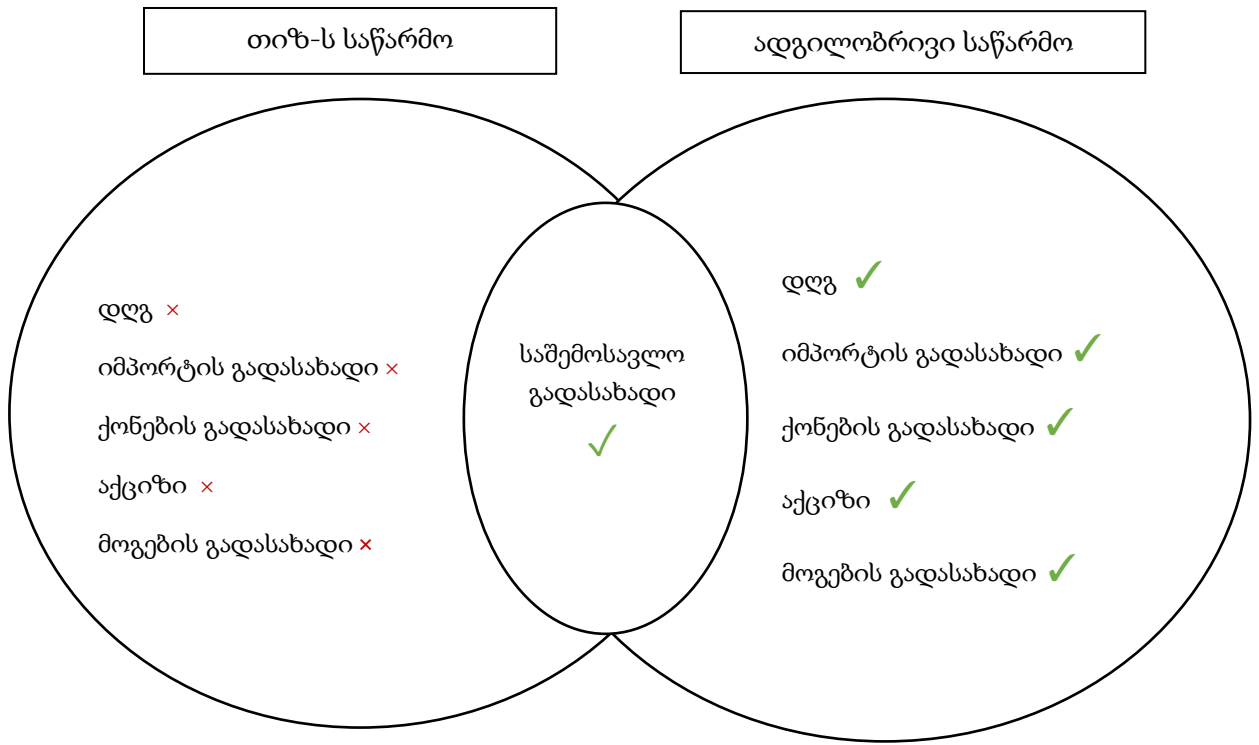

თავისუფალი ინდუსტრიული ზონების შესახებ კანონის მიხედვით, ადგილობრივი თვითმმართველობის უფლებამოსილებები არ ვრცელდება მათ საქმიანობაზე. ამან კი შეიძლება გამოიწვიოს მათი უკონტროლობა სახელმწიფოს მხრიდან. მაგალითად, აღარმოონ საეჭვო საქმიანობა, როგორიცაა ფულის გათეთრება, კორუფციული გარიგებების წარმოების ალბათობა არის მაღალი. უმჯობესი იქნებოდა თუკი უფლებამოსილებები იქნებოდა გამიჯნული. აღნიშნული არგუმენტის გასამყარებლად მოვიშველიებ 2018 წელს საინფორმაციო ანალიტიკური სააგენტოს "INFO NEWS"-ის სტატიას[14], სადაც ისინი აღწერენ თუ რა მდგომარეობა დახვდათ ფოთის თავისუფალ ინდუსტრიულ ზონაში, როდესაც ისინი დაინტერესდნენ მათი საქმიანობით და გადაწყვიტეს ადგილზე დაეთვალიერებინათ ზონა. თუმცა თიზ-ს დაცვის სამსახურმა არ მისცათ საშუალება ტერიტორიაზე შესვლის და ურჩიეს მიემართათ ხელმძღვანელობისათვის. სააგენტოს ჟურნალისტმა მიმართა ოფიციალურ მეილზე, თუმცა მათ უპასუხეს რომ თიზ-ში რეგისტრირებულ კომპანიებზე ვერ გასცემდნენ ინფორმაციას. სააგენტოს გუნდმა გადაწყვიტა

---

[14] კუპრაშვილი ვაკო. „ჟურნალისტური გამოძიება" 2018 წელი ივნისი.

https://infonews.ge/fiz/?fbclid=IwAR2ObWTo-hvm2644M81WMobv7a5sEt9zN6OOyEJv-p4TFELGmggxeYImxzc

მოეპოვებინა ინფორმაცია სხვა გზებით. რის შემდეგაც დაასკვნეს რომ ის კომპანია, რომლის შესახებაც სცადეს ინფორმაციის გამოთხოვა საერთოდ არც ფუნქციონირებდა ფოთის თავისუფალ ინდუსტრიულ ზონაში. ეს ფაქტი კი ბადებს ეჭვს რომ იქ მიმდინარეობს კომპანიათა საეჭვო საქმიანობები.

კიდევ ერთი მოსაზრება იმის შესახებ თუ რა უარყოფითი გავლენა შეიძლება იქონიოს თავისუფალი ინდუსტრიული ზონის არსებობამ ადგილობრივ საწარმოებზე. ჩემი აზრი ასეთია: თუკი თიზ-ს ტერიტორიაზე მოქმედ კომპანიებს აქვთ ამდენი შეღავათი, ფაქტია რომ წარმოება მათ ადგილობრივ კომპანიებთან შედარებით იაფი დაუჯდებათ. ისეთი კომპანიები, რომლებიც საქმიანობენ ერთი და იგივე სფეროში, ბუნებრივია არათანაბარ პირობებში მუშაობენ. ერთი და იგივე პროდუქციას თიზ-ს კომპანია უფრო იაფად აწარმოებს ვიდრე ადგილობრივი, რითაც განსხვავებული იქნება გასაყიდი ფასები და მარტივად შეუძლია გაუწიოს კონკურენცია. ეს ყველაფერი კი ვფიქრობ სტიმულს დაუკარგავს სამამულო წარმოებას.

ქვეყნის საბიუჯეტო შემოსავლები სწორედ რომ ბიზნესის განვითარებაზეა დამოკიდებული. წარმატებული ბიზნეს გარემო ეკონომიკის განვითარების საწინდარია. ქვეყანამ კი სწორედ ამ წარმატებული ბიზნესისგან უნდა მიიღოს სარგებელი. ბიზნესის წარმატება საბოლოო ჯამში მოგების სახით გვევლინება. ამ მოგებიდან წილი კი სახელმწიფოსაც ეკუთვნის და ვფიქრობ სწორედ მაღალი მოგების მქონე ბიზნესებისგან უნდა მიიღოს სახელმწიფომ შემოსავალი. თიზ-ს ტერიტორიაზე მოქმედი საწარმოების მიზანიც არის ის, რომ არსებული შეღავათების გამოყენებით გაზარდონ მათი მოგება. ეს რეალურად ასეც მოხდება და მათი მოგების მარჯა საკმაოდ დიდი იქნება. თუმცა ჩვენი ქვეყანა ამ მხრივ ვერავითარ დადებით შედეგს ვერ იღებს, რადგან ათავისუფლებს მოგების გადასახადიდან. ვფიქრობ, აღნიშნული საკითხი გადასახედია და თუნდაც ის ცვლილება, რაც ადგილობრივი საწარმოებისთვის არსებობს, გარკვეულ ეკონომიკურ ეფექტს მოიტანს.

გასაგებია რომ გადასახადების გათავისუფლებით სახელმწიფო ვერ დებულობს ეკონომიკურ ეფექტს, თუმცა არსებობს მეორე მხარეც არაპირდაპირი გადასახადების შემთხვევაში. როგორიცაა მაგალითად დღგ. დღგ-სგან გათავისუფლებული საქონელი მომხმარებლისთვის უფრო იაფია და შეიძლება მივიჩნიოთ რომ ამ გადასახადისგან გათავისუფლებით სარგებელს ნახულობს მოსახლეობის დიდი ნაწილი. თუმცა თავისუფალი ინდუსტრიული ზონების შემთხვევაში ეს დადებითი ეფექტიც არ გამოიკვეთება, რადგან ეს ზონები კანონის მიხედვით აკრძალულია საცხოვრებელი ადგილის გამოყენებისათვის. გამოდის, რომ მხოლოდ ინვესტორი იყენებს ამ შეღავათს თავის სასარგებლოდ. ეს შეღავათი ეხება კომუნალურ მომსახურებასაც. ისინი კომუნალური მომსახურების საფასურს იხდიან დღგ-ს გარეშე ანუ 18%-ით ნაკლებად. ასე მაგალითად, თბილისის თავისუფალ ინდუსტრიულ ზონაში მოქმედმა საწარმომ „BitFury", 2017 წელს ერთი თვის განმავლობაში აღნიშნული შეღავათის გამოყენებით დაზოგა 600 ათასი ლარი[15]. ფაქტია, რომ ისეთი კომპანიები, რომლებიც დიდი რაოდენობით ელექტროენერგიას მოიხმარებენ არიან ძალიან არათანაბარ პირობებში თიზ-ს საწარმოებთან შედარებით. ეს შეღავათი ადგილობრივი საწარმოების განვითარებისთვის რომ გამოიყენონ თუნდაც მხოლოდ კომუნალური გადასახადების კუთხით და მათ მისცენ განვითარების საშუალება უკეთესი ბიზნეს-გარემოს შექმნას შეუწყობენ ხელს.

მართალია, რომ თავისუფალი ინდუსტრიული ზონების შექმნის მიზანი არის ინვესტიციების მოზიდვა, თუმცა მეორე მხრივ მან შეიძლება ხელი შეუშალოს თავად სახელმწიფოს საინვესტიციო გარემოს, რადგან ინვესტორისთვის პრიორიტეტული გახდება არა სახელმწიფო, არამედ ამ სახელმწიფოში არსებული თავისუფალი ინდუსტრიული ზონა, რომლისგანაც ბუნებრივია მეტ შეღავათს მიიღებს.

გასათვალისწინებელია ისიც რომ უცხოელი ინვესტორისთვის არამხოლოდ საგადასახადო შეღავათებია მიმზიდველი არამედ გარემოც სადაც ის დააბანდებს ინვესტიციას. მაგალითად, უნდა იყოს შესაბამისი ინფრასტრუქტურული გარემო, გზა, ტრანსპორტი, კომუნიკაცია. ამ პირობების შესაქმნელად კი სახელმწიფომ უნდა გასწიოს ხარჯი. ვფიქრობ, ისეთი ტერიტორიის მოწესრიგებისათვის საიდანაც სახელმწიფო ვერ მიიღებს შესაბამის შემოსავალს, ხარჯის გაწევა არ არის გონივრული. უკეთესი იქნება ეს თანხები მიმართული იყოს ადგილობრივი საწარმოების განვითარებისა და ხელშეწყობისთვის.

---

[15] იხ.სტატია „Bitfury-სთვის დენის ტარიფი გაცილებით დაბალია, ვიდრე მოსახლეობისთვის" აბსანძე თეონა. 2018 წელი.
https://factcheck.ge/ka/story/33921-bitfury-sthvis-denis-tariphi-gatsilebith-dabalia-vidre-mosakhleobisthvis

მინდა აღვნიშნო ისიც, რომ თიზის საწარმოების ინტერესი ადგილობრივ საწარმოებთან თანამშრომლობისა ნაკლებია, გამომდინარე იქიდან რომ საგადასახადო კოდექსის თანახმად, საქართველოს კანონმდებლობით რეგისტრირებული პირისგან საქონლის შეძენის შემთხვევაში უნდა გადაიხადოს საქონლის საბაზრო ფასის 4%. ბუნებრივია, ის ამჯობინებს საქონლის იმპორტს სხვა ქვეყნისგან რადგან გათავისუფლებულია როგორც იმპორტის გადასახადისგან ასევე დღგ-სგან. ეს არის კიდევ ერთი მაგალითი იმისა რომ ადგილობრივ ბიზნეს-სუბიექტებს მათი საქმიანობით არ აქვთ არანაირი სარგებელი.

მაგალითისთვის შევქმნათ საწარმოს ბალანსი, იმისათვის რომ დავინახოთ თუ როგორ შეიძლება შეიცვალოს ადგილობრივი კომპანიის მოგება იმ პირობების შემთხვევაში, რომლებიც აქვთ თავისუფალ ინდუსტრიულ ზონებში რეგისტრირებულ საწარმოებს. ვთქვათ გვყავს კომპანია "X". განვიხილოთ ორი შემთხვევა როცა ის საქმიანობს თიზ-ს ტერიტორიაზე და მის გარეთ.

ცხრილი N7: კომპანიის ბალანსი თიზ-ის გარეთ და თიზ-ში.[16]

| კომპანიის ბალანსი | თიზ-ს გარეთ | თიზ-ში | საგადასახადო შეღავათი |
|---|---|---|---|
| ბრუნვა | 2 000 000 | 2 000 000 | |
| საქონლის ღირებულება იმპორტირებული საქონელი იმპორტის გადასახადი სხვა ხარჯები | 765 000 300 000 15 000 450 000 | 750 000 300 000 0 450000 | 15 000 |
| საოპერაციო ხარჯები ქონების გადასახადი სხვა საოპერაციო ხარჯი | 615 000 15 000 600 000 | 600 000 0 600 000 | 15 000 |
| საპროცენტო გადასახადი და ცვეთა | 150 000 | 150 000 | |
| მოგების გადასახადის გადახდამდე წმინდა შემოსავალი | 470 000 | 500 000 | |
| მოგების გადასახადი | 70 500 | 0 | 70 500 |
| მოგება გადასახადის გადახდის შემდეგ | 399 500 | 500 00 | |

როგორც ცხრილიდან ჩანს, საგადასახადო შეღავათები მოდის იმპორტის გადასახადზე, ქონების გადასახადზე და მოგების გადასახადზე. იმპორტის გადასახადის განაკვეთად სამუალოდ ავიღეთ 5% რადგან საქართველოში საქონლის სახეობიდან გამომდინარე პროცენტი არის განსხვავებული. კერძოდ, მოქმედებს 0%, 5% და 12% იმპორტის გადასახადის განაკვეთები. როგორც ჩანს, კომპანია ზოგავს საგადასახადო შეღავათებით მნიშნელოვან ფინანსურ რესურსს 100 500 ლარს. აქვე გასათვალისწინებელია თავისუფალ ინდუსტრიულ ზონაში მოქმედი საწარმოებისთვის მოსაკრებელი, რომელმაც შესაძლოა შეადგინოს აღნიშნული თანხის 50%. ამ შემთხვევაში კომპანიის მთლიანი წმინდა დანაზოგი იქნება 50 250 ლარი. საბოლოო ჯამში თიზი-ს გარეთ მოქმედი საწარმო დაზოგავდა თავისი წმინდა მოგების 12,5%-ს რომ მიეღო იგივე შეღავათები რაც აქვთ თიზი-ს ტერიტორიაზე არსებულ საწარმოებს.

---
[16] ცხრილი შედგენილია ავტორის მიერ მოცემული მონაცემების მიხედვით.

### *3.2 საერთაშორისო დაბეგვრის უცხოური გამოცდილება და მათი შეფასება*

როგორც ეროვნული საწარმოები, ისე საერთაშორისო საწარმოებიც მიისწრაფვიან მოგების მაქსიმიზაციისკენ. თუმცა ისინი უფრო რთულ პირობებში მუშაობენ, რადგან მათ უწევთ როგორც ქვეყნის კანონმდებლობის დაცვა ასევე იმ ქვეყნებისა, რომლებშიც ისინი ეწევიან საქმიანობას. ისინი ცდილობენ შეიცირონ საგადასახადო წნეხი. ამ მიზეზით იყენებენ ორ ძირითად მიდგომას: ტრანსფერტული ფასწარმოქმნა და საგადასახადო სამოთხე.

ტრანსფერტული ფასწარმოქმნა გულისხმობს საქონელსა და მომსახურებაზე ფასების დაწესებას სათავო ფირმის ერთი ფილიალის ან შვილობილი საწარმოს მიერ მეორე ფილიალისთვის. იგი გამოიანგარიშება საბაზრო მეთოდით, რომელიც გაიანგარიშება ღია ბაზრის ფასის საფუძველზე და არასაბაზისო მეთოდით, რომელიც დგინდება მყიდველისა და გამყიდველის მოლაპარაკების საფუძველზე საქონლის თვითღირებულებაზე დაყრდნობით. საერთაშორისო საწარმოები ძირითადად იყენებენ არასაბაზისო ტრანსფერტულ ფასებს. მას გააჩნია როგორც დადებითი ასევე უარყოფითი მხარეები, თუმცა ის შეიძლება ხელსაყრელი იყოს საერთაშორისო საწარმოებისთვის. კერძოდ, მისი საშუალებით საწარმომ შეიძლება შეამციროს გადასახადის საერთო სიდიდე. მაგალითად, თუ კომპანია მუშაობს ორ ქვეყანაში, ერთში კორპორაციულ მოგებაზე გადასახადი მაღალია მეორეში კი დაბალი, მას შეუძლია გაზარდოს ტრანსფერტული შვილობილი კომპანიისთვის დადგენილი ფასები მაღალგადასახადიანი ქვეყნიდან და შეამციროს ტრანსფერტული ფასები შვილობილი კომპანიისათვის დაბალგადასახადიანი ქვეყნიდან. ამის საშუალებით ის ამცირებს მოგების სიდიდეს პირველი შვილობილი კომპანიისათვის და ზრდის მას მეორე კომპანიისათვის. შედეგად, სათავო კომპანიის მოგება გადადის მაღალსაგადასახადოგანაკვეთიანი ქვეყნიდან დაბალ საგადასახადოგანაკვეთიან ქვეყანაში და საერთო საგადასახადო წნეხი მცირდება. ამგვარად, ტრანსფერტული ფასწარმოქნის მეშვეობით საერთაშორისო კომპანიები ხშირად ახდენენ საგადასახადო წნეხის გადატანა-გადანაწილებას.

ამ ტექნოლოგიებში ისეთი სახელმწიფო როგორიცაა აშშ კარგად არის გათვიცნობიერებული და მნიშვნელოვან ყურადღებას უთმობენ საფასო პოლიტიკას. საგადასახადო ორგანოები აკონტროლებენ კომპანიებს რათა თავი არ აარიდონ საგადასახადო ვალდებულებებს. ისინი ამოწმებენ ფირმებს შორის მოლაპარაკების შედეგად დადებულ ფასს, რის გამოც ხშირ შემთხვევაში მათ შორის წარმოიქმნება კონფლიქტები.

განსხვავებული მოსაზრებები არსებობს ასევე საგადასახადო სამოთხესთან დაკავშირებით. საგადასახადო სამოთხე და ტრანსფერტული ფასწარმოქნა ხელსაყრელია საერთაშორისო საწარმოების აქციონერებისათვის, მეორე მხრივ კი ის შეიძლება ნაკლებად სასარგებლო აღმოჩნდეს სათავო ორგანიზაციის ქვეყნის საბიუჯეტო შემოსავლებზე, რადგან ადგილი აქვს მათ შემცირებას. მაშინ როცა ამ შემოსავლებით შესაძლებელია მნიშვნელოვანი სოციალური პრობლემების გადაწყვეტა. ეს ნიშნავს რომ საერთაშორისო კომპანია სამართლიანად არ იხდის გადასახდს და გადასახადის სხვა გადამხდელები მიერ ხდება პრობლემების გადაჭრა.

საგადასახადო სამოთხის ქვეყნების ეკონომიკა შეიძლება იყოს მაღალგანვითარებული. მაგალითად, კაიმანის კუნძულებზე, მისი წარმატება განპირობებულია საერთაშორისო კომპანიებისათვის გაწეული მომსახურების მაღალი ხარისხით, ეკონომიკის სიძლიერით, რაც განპირობებული მთავრობის სტაბილურობით, მისი ეკონომიკის სიძლიერე განპირობებულია ტურიზმის, საბანკოს სისტემისა და ოფშორული კორპორაციების პოპულარობით. სხვა ოფშორული ზონებისაგან განსხვავებით კაიმანის კუნძულები მიჰყვება საერთაშორისო საგადასახადო რეგულაციებს.[17] სანაცვლოდ, იქმნება სამუშაო ადგილები მაღალი ანაზღაურებით პროფესიონალებისათვის, ყვავის ტურისტული ბიზნესი და ქვეყანა მოსახლეობის ცხოვრების დონით ერთ-ერთი მოწინავე მსოფლიოში. თუმცა, მეორე მხრივ, სხვა ქვეყნების საგადასახადო ორგანოებს უქმნის საკმაოდ პრობლემებს, გადასახადების თავიდან აცილების შესაძლებლობის მიცემით.

---
[17] კაიმანის კუნძულები კომპანია რეგისტრაცია.

უცხოური მოგების დაბეგვრა სხვადასხვა სახელწიფოში არაერთგვაროვანია. განვიხილოთ აშშ-ს მაგალითი. ქვეყანაში უცხოური შემოსავალი გენერირდება სამი წყაროდან: პროდუქციის ან მომსახურების ექსპორტი, უცხოური ფილიალების საქმიანობა და უცხოური შვილობილი საწარმოების საქმიანობა.

აშშ-ში საქონლისა და მომსახურების ექსპორტიდან მიღებული მოგება განიხილება ისე როგორც შიგა ბაზარზე მიღებული მოგება. ამერიკული საგადასახადო კოდექსი სტიმულირების მიზნით დიდი ხნის განმავლობაში ნებას რთავდა საგარეო-სავაჭრო კორპორაციების შექმნას. მათ, მათ თავიანთი საქმიანობის მნიშვნელოვანი ნაწილი უნდა განეხორციელებინა საზღვარგარეთ. ფირმას, რომელიც შეასრულებდა კორპორაციის ამ პირობებს შეეძლო შეემცირებინა თავისი ფედერელური გადასახადი საექსპორტო საქმიანობიდან. ვაჭრობის მსოფლიო ორგანიზაციამ გააკეთა განმარტება რომ საგარეო-სავაჭრო კორპორაციისადმი საგადასახადო შედავათების გაწევა არ ეწინააღმდეგებოდა ექსპორტის არაკეთილსინდისიერ სუბსიდირებას. ამიტომ, აშშ-ს კონგრესმა მიიღო კანონი დაბეგვრიდან მოგების ექსტერიტორიული გამორიცხვის შესახებ, რომელიც მსოფლიო სავაჭრო ორგანიზაციამ ჩათვალა მისი წესდების დარღვევად. ევროკავშირი კი იტოვებდა უფლებას აშშ-დან მიღებულ საქონელსა და მომსახურებაზე დაედო ტარიფები სანამ აშშ არ თავის საგადასახადო კოდექსს მსოფლიო სავაჭრო ორგანიზაციის ნორმებთან შესაბამისობაში არ მოიყვანდა. თუმცა, ევროკავშირი არ მიმართავს აგრესიულ ზომებს რათა თავისუფალი ვაჭრობისთვის ხელი არ შეეშალა, ამასთან მისთვის ამერიკული ბაზარი ძალიან მნიშვნელოვანია.

რაც შეეხება უცხოური კომპანიების ფილიალების დაბეგვრას, ამერიკაში არსებობს შემდეგი პრაქტიკა: (Гриффин, 2006: 958-962) მათ არ გააჩნიათ იურიდიული პირის უფლება, მათი მოგება სამართლებრივი კუთხით სათავო კომპანიის მოგების შემადგენელი ნაწილია, აქედან გამომდინარე უცხოური ფილიალების შემოსავლების

---

https://www.offshorecompany.com/ka/company/cayman-islands/

ზრდა იწვევს სათავო კომპანიის დასაბეგრ ბაზას, მიუხედავად იმისა, ხდება თუ არა მათი შემოსავლების რეპატრიაცია სათავო კომპანიის ქვეყანაში.

განსხვავებული მიდგომაა შვილობილი კომპანიების მიმართ. მათ გააჩნიათ დამოუკიდებელი იურიდიული პირის სტატუსი. თუ შვილობილი კომპანიის მოგების რეინვესტირება ხდება იმავე შვილობილ კომპანიაში აშშ-ს სათავო კომპანია არაა ვალდებული თავის შემოსავალში ასახოს ეს მოგება. ამ შემთხვევაში მოქმედებს განვადების წესი, რაც გულისხმობს ასეთი შემოსავლების დაბეგვრას იმ შემთხვევაში თუკი სათავო კომპანია დივიდენდების სახით დაიბრუნებს მას. ეს წესი ასტიმულირებს აშშ-ს ფირმების საერთაშორისო საქმიანობას, მისი საშუალებით კომპანიების ახორციელებენ დიდი მოცულობით საგადასახადო სახსრების ეკონომიას, რაც საშუალებას აძლევთ ახალ ბაზრებში შეღწევისა და თავიანთი საქმიანობის განვითარებისა.

აშშ საგადასახადო კოდექსის მიხედვით, უცხოური კორპორაციების საგადასახადო მოგებას ყოფს აქტიურ ანუ ტრადიციული კომერციული საქმიანობიდან (წარმოება, მარკეტინგი) მიღებულ მოგება და პასიურ ანუ პასიური ქმედებებით(დივიდენდები, პროცენტები კრედიტებზე და ა.შ.) მიღებულ მოგებად.

ამერიკულ ფირმებს შეუძლიათ გადაავადონ მათ მიერ კონტროლირებადი უცხოური კორპორაციების მიერ მიღებული აქტიური მოგება ხოლო პასიური არ ექვემდებარება გადავადების წესს. აშშ-ს ხელისუფლება ამ გზით მკაფიოდ აცალკევებს თავიანთი ფირმების საერთაშორისო საქმიანი აქტივობისა და ამერიკული გადასახებიდან თავის არიდების შესაძლებლობის შეზღუდვას.

კიდევ ერთი მეთოდი რომელსაც იყენებს აშშ ეს არის საგადასახადო კრედიტი, აშშ საგადასახადო კოდექსი ხშირ შემთხვევაში საშუალებას აძლევს ამერიკულ ფირმებს შეამცირონ კორპორაციული მოგების გადასახადი იმ სიდიდით, რა სიდიდის ამავე გადასახადს იხდიან უცხოური ფილიალები და შვილობილი კომპანიები. თუმცა, უცხოურ მოგებაზე საგადასახადო დათმობა არ უნდა აღემატებოდეს გადასახადის სიდიდეს, რომლითაც აშშ-ში იბეგრება უცხოური საქმიანობა. საგადასახადი კრედიტი ვრცელდება მხოლოდ კორპორაციული მოგების გადასახადზე და მისი გამოყენება სხვა სახის გადასახადზე დაუშვებელია.

აშშ-ს გარკვეული საგადასახადო შეღავათების მიზნით გაფორმებული აქვს ხელშეკრულებები 60 ქვეყანასთან, სადაც განსაზღვრულია გადასახადების შემცირების პირობები. ეს ხელშეკრულებები არის ორმხრივი და ნიშნავს რომ ქვეყანა სხვა ქვეყნის ფირმებს სთავაზობს შეღავათიანი დაბეგვრის პირობებს მხოლოდ მაშინ თუკი სხვა ქვეყანა ანალოგიურ პირობებს შეუქმნის ამ ქვეყანას.

საერთაშორისო საგადასახადო კონფლიქტების ერთ-ერთ წყაროს ასევე წარმოადგენს უცხოური ფირმების შესახებ ადგილობრივი პოლიტიკოსების შეხედულებები. ისინი თვლიან რომ ტრანსფერული ფასების საშუალებით ასეთი ფირმები თავს არიდებენ გადასახადებს. ასე მაგალითად, იაპონია მკაცრად ამოწმებს უცხოური ფირმების ფასწარმოქმნის პოლიტიკას რის გამოც ხშირად აქვს პრეტენზიები უცხოური კომპანიების მიმართ. მან ამ მიზნით გამკაცრა აუდიტორული სამსახურების მუშაობა.

განხვავებით ჩვენი ქვეყნისა, თავისუფალი ინდუსტრიული ზონები მრავალ ქვეყანაში მოქმედებს წარმატებით და ჩვენს ქვეყანას ასე ვითქვათ, ამ მხრივ ჰყავს კონკურენტები. შორს რომ არ წავიდეთ, ჩვენს მეზობელ თურქეთში აღნიშნული ზონები წარმატებით ფუნქციონირებს. სულ არსებობს 21 თავისუფალი ინდუსტრიული ზონა, სადაც მოქმედებს შემდეგი შეღავათები:[18]

- დაბალი საშემოსავლო და მოგების გადასახადი;
- გათავისუფლებულია დღგ-სა და სხვა მოხმარებაზე გადასახადისაგან;
- უცხოური ვალუტის გამოყენების შესაძლებლობა;
- საბაჟო გადასახადისგან გათავისუფლებულია იმპორტირებული საქონელი;
- თურქეთის პორტებთან ადვილი წვდომა;
- საბანკო ტრანზაქციების გადასახადებისგან გათავისუფლება;
- არ არის შეზღუდვა საქონლის ფასსა და ხარისხზე;
- უცხოური კაპიტალის დაბანდება შესაძლებელია განუსაზღვრელი რაოდენობით;
- თანამედროვე, მიმზიდველი და მრავალფეროვანი ინფრასტრუქტურული ბიზნეს-გარემო;

---

[18] Turkey Free Trade Zones

https://www.systemday.com/turkey/turkey-free-trade-zones.php

- აკრძალულია გაფიცვები დასაქმებულების მხრიდან, ასევე დამსაქმებლის მხრიდან დასაქმებულის სამსახურიდან დათხოვა დასაბუთებული მიზეზების გარეშე;
- არ არის შეზღუდვა დროზე თუ რამდენი ხანი დარჩება საქონელი თავისუფალი ინდუსტრიული ზონის ტერიტორიაზე;
- მუშა ხელისა და მიწის სიიაფე თავისუფალი ინდუსტრიული ზონის ტერიტორიაზე;
- უპირატესობით სარგებლობენ ადგილობრივ ბაზრებზე.

მოცემულ ზონებში ხორციელდება სხვადასხვა სახის საქმიანობა: მშენებლობა, IT მომსახურება, დაზღვევა, სასაწყობო მომსახურება, ტვირთის ჩატვირთვა-გადმოტვირთვა, საკონსულტაციო მომსახურება, 24 საათიანი პირადი უსაფრთხოება, საწარმოო ობიექტების გაქირავება და ა.შ. თიზ-ის ტერიტორიაზე საწარმოები უზრუნველყოფილი არიან ტექნიკურ-ინფრასტრუქტურული მომსახურებით. როგორიცაა: ელექტროენერგია, გაზი, წყალი, კომუნიკაცია. ამ ყველაფერზე ზრუნავს სახელმწიფო, რათა უფრო მიმზიდველი გახადოს უცხოელი ინვეტორისთვის ეს ტერიტორიები.

აღსანიშნავია რომ არის გარკვეული მსგავსებები ჩვენს ქვეყანასთან მიმართებაში საგადასახადო შეღავათების კუთხით, თუმცა ვფიქრობ, საქართველო ჯერჯერობით არ არის მზად აღნიშნული სფეროს გასავითარებლად, მას სჭირდება პირველ რიგში ადგილობრივი ეკონომიკის წინსვლა და უმჯობესი იქნება თუკი ამ კუთხით განაგრძობს ის მიმზიდველი ბიზნეს-გარემოს შექნისთვის პირობების დაწესებას.

დასკვნა

ყოველივე ზემოთქმულიდან გამომდინარე, შეგვიძლია დავასკვნათ, რომ საერთაშორისო დაბეგვრას მნიშვნელოვანი როლი აქვს ადგილობრივი ეკონომიკის განვითარებასა და მის დაცვაში უცხოელი ინვესტორებისაგან, კონკურენციის პირობებიდან გამომდინარე.

ნაშრომის შესრულებამ, საშუალება მოგვცა გავცნობოდით საერთაშორისო დაბეგვრის თეორიულ-მეთოდოლოგიურ საკითხებს, წარმოდგენა გვქონდა საგადასახადო კუთხით ჩვენს ქვეყანაში არსებულ პრობლემებზე და ამასთან დაკავშირებით ჩამოგვეყალიბებინა საკუთარი მოსაზრება.

საქართველო მაქსიმალურად ცდილობს ჩაერთოს გლობალურ პროცესებში, განსაკუთრებით საერთაშორისო ვაჭრობის კუთხით, ამიტომ საერთაშორისო დაბეგვრის საკითხი დღითიდღე აქტიური ხდება. ქვეყანა ცდილობს საერთაშრისო ხელშეკრულებებით დაამყაროს სავაჭრო ურთიერთობები სხვა ქვეყნებთან, შესთავაზოს მათ გარკვეული შეღავათები, რომლისგან სარგებელს ნახავს, როგორც თავად, ასევე ხელშემკვრელი სახელმწიფო.

ნაშრომზე მუშაობამ მიგვიყვანა იმ დასკვნამდე, რომ საერთაშორისო დაბეგვრის პროცესებში უპირველესად უნდა იყოს გათვალისწინებული ადგილობრივი ეკონომიკის როლი, ბიზნეს-სუბიექტების საქმიანობა და ის შესაძლო შედეგები რაც შეიძლება მოჰყვეს უცხოელ ინვესტორთა მოზიდვას. ზედმეტად ლიბერალური საგადასახადო პოლიტიკაც არ იძლევა შესაბამის ფისკალურ ეფექტს, კონკურენციას უწევს ადგილობრივ საწარმოებს და საბოლოო ჯამში შესაძლოა მივიღოთ უფრო მეტი უარყოფითი შედეგი ვიდრე დადებითი.

ამ პრობლემების აღმოფხვრისათვის მნიშვნელოვანია ასევე საგადასახადო კანონმდებლობის სრულყოფა და ადგილობრივი ბიზნეს-სუბიექტების წინა პლანზე წამოწევის საკითხი. ქვეყანა ჯერ ადგილობრივი ეკონომიკით უნდა იყოს ძლიერი, რათა შეძლოს უცხოელი ინვეტორებისაგან სარგებლის მიღება და განხორციელებული პროექტების თავის სასარგებლოდ გამოყენება. თუ ისინი უპირატეს მდგომარეობაში აღმოჩნდებიან, სამამულო წარმოება ვერ განვითარდება. მნიშვნელოვანია ასევე უცხოური გამოცდილების გაზიარება, თუმცა, უნდა იყოს გათვალისწინებული ჩვენი ქვეყნის თავისებურებები და პირდაპირ სხვა ქვეყნის გამოცდილების გადმოტანა არ იქნება გონივრული გარკვეული შენიშვნების გათვალისწინების გარეშე.

უმჯობესია თუკი მხოლოდ თეორიულ ასპექტებს არ დავეყრდნობით და განხორციელებული ღონისძიებების შესახებ დასკვნებს გავაკეთებთ რეალობიდან გამომდინარე. ვგულისხმობ, თავისუფალი ინდუსტრიული ზონების მუშაობის გადახედვას. თეორიულად ეს ყველაფერი კარგად ჟღერს, თუმცა რამდენად ამართლებს მოლოდინი და ჩვენი ქვეყანა რამდენად არის მზად მსგავსი ზონების მუშაობის ხელშეწყობისთვის ეს უკვე ცალკე საკითხია. ამის შესახებ ნაშრომში გვაქვს რამდენიმე მაგალითი.

საბოლოო ჯამში შეგვიძლია ვთქვათ, რომ პირველ რიგში უნდა იქნას გაცნობიერებული ქვეყნის საჭიროებები, ეკონომიკის ამოცანები და მიზნები და საგადასახადო პოლიტიკის მთავარ მიმართულებად უნდა იქნეს არჩეული ადგილობრივი ეკონომიკის განვითარება, პრიორიტეტული სფეროები, რომლებიც შემდგომში უზრუნველყოფენ ქვეყნის ეკონომიკურ სტაბილურობას.

## გამოყენებული ლიტერატურა

1. აბსანძე, თ. (2018). BitFury-სთვის დენის ტარიფი გაცილებით დაბალია, ვიდრე მოსახლეობისთვის. (მოძიებულია 2 მაისი, 2020 წელი). https://factcheck.ge/ka/story/33921-bitfury-sthvis-denis-tariphi-gatsilebith-dabalia-vidre-mosakhleobisthvis

2. აბსანძე, თ. (2017). რა აფერხებს საქართველოში თიზების განვითარებას? (მოძიებულია 2 მაისი, 2020). https://forbes.ge/news/3229/ra-aferxebs-Tizebis-ganviTarebas?fbclid=IwAR1KzOADRzjBRswY0wcMzrTrLQiUPy5ItrjEIMvBRlFqczSA3fa3M5GxigQ

3. ვერულიძე, ვ. (2015). გადასახადები და საგადასახადო დაბეგვრა. გამომცემლობა ბათუმის შოთა რუსთაველის სახელმწიფო უნივერსიტეტი , ბათუმი. 2015.

4. კაიმანის კუნძულები კომპანია რეგისტრაცია https://www.offshorecompany.com/ka/company/cayman-islands/

5. კუპრაშვილი, ვ. (2018). ჟურნალისტური გამოძიება. (მოძიებულია 5 მაისი, 2020). https://infonews.ge/fiz/?fbclid=IwAR2ObWTo-hvm2644M81WMobv7a5sEt9zN6OOyEJv-p4TFELGmggxeYImxzc

6. ყუფარაძე, გ. (2013). საგადასახადო პოლიტიკა და პირდაპირი უცხოური ინვესტიციები საქართველოში. მონოგრაფია თბილისი. https://library.iliauni.edu.ge/wp-content/uploads/2017/03/giorgi-qhupharadze-sagadasakhado-politika-da-pirdapiri-utskhouri-investitsiebi-saqarthveloshi.pdf

7. „საქართველოს კანონი თავისუფალი ინდუსტრიული ზონების შესახებ" ; https://matsne.gov.ge/ka/document/view/21994?publication=8

8. ,,საქართველოს საგადასახადო კოდექსი" (მუხლი 124-134); https://matsne.gov.ge/ka/document/view/1043717

9. საქართველოს სტატისტიკის ეროვნული სამსახური; https://www.geostat.ge/ka/modules/categories/191/pirdapiri-utskhouri-investitsiebi

10. საქართველოს ფინანსთა სამინისტრო: https://mof.ge/5127

11. საქართველოს პარლამენტი, (2018). საქართველოს თავისუფალი ინდუსტრიული ზონები.
http://parliament.ge/ge/ajax/downloadFile/98813/%E1%83%A1%E1%83%90%E1%83%A5%E1%83%90%E1%83%A0%E1%83%97%E1%83%95%E1%83%94%E1%83%9A%E1%83%9D%E1%83%A1_%E1%83%97%E1%83%90%E1%83%95%E1%83%98%E1%83%A1%E1%83%A3%E1%83%A4%E1%83%90%E1%83%9A%E1%83%98_%E1%83%98%E1%83%9C%E1%83%93%E1%83%A3%E1%83%A1%E1%83%A2%E1%83%A0%E1%83%98%E1%83%A3%E1%83%9A%E1%83%98_%E1%83%96%E1%83%9D%E1%83%9C%E1%83%94%E1%83%91%E1%83%98?fbclid=IwAR04WdxIDUg4BHXuZR7jD_zp5apiy-52F3XJfvTggPdctKJHbB2gSRThFos

12. საქართველოს კანონი „მეწარმეთა შესახებ".
https://matsne.gov.ge/ka/document/view/28408?publication=64

13. ურიდია, გ. (2013). საერთაშორისო დაბეგვრა და ტრანსფერული ფასწარმოქმნა.
http://www.aaf.ge/index.php?menu=1&jurn=0&rubr=6&mas=1963

14. Гриффин Р., Пастей М. (2006). Международный бизнес, 4-ое изд. Питер.- 2006.

15. Полежарова, Л. В. (2009). Международное двойное налогообложение: механизм устранения в Российской Федерации. Москва.

16. Abuselidze, G. (2005a). The prospects of modern budget revenue in the aspect of the new Tax Code. Journal of Social Economy, 1, 49-60.

17. Abuselidze, G. (2005b). To optimize legal foundations of the budget system. Journal of Social Economy, 4, 72-80.

18. Abuselidze, G. (2008). Taxation heaviness in Georgia and its optimization–for tax saving hints. Economics and Business, 4, 74-82.

19. Abuselidze, G. (2009a). About inevitability of budgetary code receiving for fiscal politics. Economics and Business, 1, 42-46.

20. Abuselidze, G. (2009b). Tax-payment infringement and questions of protection of the Tax-bearers` rights. Journal of Economics and Business, 4, 81-92.

21. Abuselidze, G. (2012b). Optimal Tax Burden and Budget System-the Factor of Macroeconomical Stabilization. Journal of Applied Finance and Banking, 2(6), 121-130.

22. Abuselidze, G. (2012c). The Influence of Optimal Tax Burden on Economic Activity and Production Capacity. Intellectual Economics, 6(4), 493-503.
23. Abuselidze, G. (2013b). Optimal Tax Policy–Financial Crisis Overcoming Factor. Asian Economic and Financial Review, 3(11), 1451-1459.
24. Abuselidze, G. (2015). Formation of Tax Policy in the Aspect of the Optimal Tax Burden. Abuselidze, George. International Review of Management and Business Research, 4, 601-610.
25. Abuselidze, G. (2018b). Optimal Fiscal Policy–Factors for the Formation of the Optimal Economic and Social Models. J. Bus. Econ. Review, 3(1), 18-27.
26. Abuselidze, G. (2018c). Fiscal policy directions of small enterprises and anti-crisis measures on modern stage: During the transformation of Georgia to the EU. Science and studies of accounting and finance: problems and perspectives, 12(1), 1-11. doi:10.15544/ssaf.2018.01
27. Abuselidze, G. (2019a). Analysis of the Formation and Use of Budgetary Policies Ensuring the Socio-Economic Development of Territorial Units. Economic Archive, 2, 3-15.
28. Company functions and company documents, მოძიებულია 7 მაისი, 2020. https://www.systemday.com/turkey/turkey-free-trade-zones.php
29. European Bank Brochure – „Investing for change"., მოძიებულია 29 მარტი.2020. file:///C:/Users/user/Downloads/georgia-brochure%20(1).pdf
30. The world Bank. https://www.doingbusiness.org/en/about-us